\documentclass[USenglish,oneside,twocolumn]{article}

\usepackage[utf8]{inputenc}%(only for the pdftex engine)
\usepackage[big]{dgruyter_NEW}
 
\usepackage{amsthm}
\usepackage{enumitem}
\newcommand{\remove}[1]{}
\newcommand{\ignore}[1]{}
\newcommand{\pie}[2]{%
\begin{tikzpicture}
 \draw (0,0) circle (1ex);\fill (1ex,0) arc (#1:#2:1ex) -- (0,0) -- cycle;
\end{tikzpicture}%
}

\newcommand{\fullc}{\pie{0}{360}}
\newcommand{\halfc}{\pie{0}{180}}
\newcommand{\emptyc}{\pie{0}{0}}

\def\OP{\mathsf{Ops}}
\def\PP{\mathsf{PreProcessor}}
\def\Auth{\mathsf{Auth}}
\def\IndexOP{\mathsf{Index\_OP}}

\newcommand{\db}{{\mathrm{DB}}}
\newcommand{\DB}{\db}

\newcommand{\EncDB}{\mathsf{EncDB}}
\newcommand{\EDB}{\mathrm{EDB}}

\newcommand{\encres}{\mathrm{encres}}
\newcommand{\Init}{\mathsf{Init}}
\def\Query{\mathsf{Query}}
\def\op{\mathsf{op}}

\def\CompleteClientAuth{\mathsf{CompleteClientAuth}}

\def\TokenTransfer{\mathsf{TokenTransfer}}
\def\KeyTransfer{\mathsf{KeyTransfer}}
\def\KeyReceive{\mathsf{KeyReceive}}

\def\QueryPreProcessing{\mathsf{QueryPreProcessing}}
\def\FileRead{\mathsf{FileRead}}
\def\FileWrite{\mathsf{FileWrite}}

\def\sessk{\mathsf{sessk}}

\def\query{\mathsf{query}}
\def\encquery{\mathsf{encquery}}

\def\inp{\mathsf{input}}

\newtheorem{claim}{Claim}[section]

\newcommand{\secparam}{\lambda}
\newcommand{\secp}{\secparam}

\def\cL{{\cal L}}
\def\cM{{\cal M}}

\def\cQ{{\cal Q}}

\def\cS{{\cal S}}

\def\cq{{\cal q}}
\def\negl{{\rm negl}}

\newcommand{\out}{{\mathsf{out}}}
\newcommand{\ID}{\mathsf{ID}}

\newcommand{\KeyGen}{\mathsf{KeyGen}}

\newcommand{\Sim}{\mathsf{Sim}}

\def\ShowAuthNotes{0}

\ifnum\ShowAuthNotes=1

\newcommand{\authnote}[2]{\textcolor{red}{\parbox{0.9\linewidth}{[{\footnotesize {\bf #1:} { {#2}}}]}}}

\else
\newcommand{\authnote}[2]{}
\fi

\newcommand{\dnote}[1]{\authnote{Dnote}{#1}}

\newcommand{\Adv}{\mathsf{Adv}}

\newcommand{\absnewline}{\ifnum\full=0 \\ \fi}

\newcommand{\key}{\mathsf{K}}

\newcommand{\program}{\mathsf{P}}

\newcommand{\HW}{\mathsf{HW}}
\newcommand{\Load}{\mathsf{Load}}
\newcommand{\Execute}{\mathsf{Execute}}
\newcommand{\Enclave}{\mathsf{E}}

\newcommand{\msg}{\mathsf{msg}}

\newcommand{\Real}{\mathsf{Real}}
\newcommand{\Ideal}{\mathsf{Ideal}}
\providecommand{\Auth}{\mathsf{Auth}}

\usepackage{tikz}
\usepackage{pgfplots}
\usepackage{float}
\usepackage{subcaption}

\def\full{0} %% set to 0 for springer proceedings

\def\name{StealthDB}
  
\begin{document}
 \author*[1]{Dhinakaran Vinayagamurthy}

  \author[2]{Alexey Gribov}

  \author[3]{Sergey Gorbunov}

  \affil[1]{IBM Research India, E-mail: dvinaya1@in.ibm.com. Work done while at University of Waterloo.}

  \affil[2]{Symbiont.io, E-mail: gribov.alesha@gmail.com. Work done while at Stealthmine Inc.}

  \affil[3]{University of Waterloo and Algorand, E-mail: sgorbunov@uwaterloo.ca}

     \title{\huge \name{}: a Scalable Encrypted Database with Full SQL Query Support}

  \runningtitle{\name{}}

\begin{abstract}
{Encrypted database systems provide a great method for 
protecting sensitive data in untrusted infrastructures. 
These systems are built using either special-purpose cryptographic algorithms that support operations over encrypted data, or 
by leveraging trusted computing co-processors. 
Strong cryptographic algorithms (e.g., public-key encryptions, garbled circuits) usually result in high performance overheads, while weaker algorithms (e.g., order-preserving encryption) result in large leakage profiles. 
On the other hand, some encrypted database systems (e.g., Cipherbase, TrustedDB)
leverage non-standard 
trusted computing devices, and are designed to work around 
the architectural limitations of the specific devices used. \\
In this work we build StealthDB -- an encrypted database system from Intel SGX.
Our system can run on any newer generation Intel CPU. 
StealthDB has a very small trusted computing base, scales to large transactional workloads, requires minor DBMS changes, and provides a relatively strong security guarantees at steady state and during query execution. Our prototype on top of Postgres supports the full TPC-C benchmark with a 30\% decrease in the average throughput over an unmodified version of Postgres operating on a 2GB unencrypted dataset.
%\dnote{what's the performance of the system? either say it also provides better performance, or say that the trade-off with the above advantages is a small performance loss.}
}\end{abstract}

  \keywords{Encrypted databases, Intel SGX}
%  \classification[PACS]{}
 % \communicated{...}
 % \dedication{...}

  \journalname{Proceedings on Privacy Enhancing Technologies}
%\DOI{Editor to enter DOI}
%  \startpage{1}
%  \received{..}
%  \revised{..}
%  \accepted{..}
%
%  \journalyear{..}
%  \journalvolume{..}
%  \journalissue{..}

\maketitle

%\thispagestyle{empty}
%\newpage

%\pagenumbering{arabic}
%\setcounter{page}{1}

\section{Introduction}

%Over the last decade, IT infrastructure has been undergoing major changes.
%%Today the IT infrastructure is undergoing a major revolution. 
%Classically, enterprise data was held and processed within a company's 
%data center. Today, more and more companies are moving their data to 
Over the last decade, storing and processing of enterprise data for a lot of companies has moved from the company's data center to 
third party public cloud infrastructure or service providers like AWS, Microsoft Azure and Google Cloud. 
These infrastructures are operated and maintained by potentially untrusted operators.
Also, the infrastructure is shared between numerous clients. For instance, a single AWS physical instance may 
co-locate a number of virtual client instances. 
Given these \textit{features}, protecting the confidentiality and integrity of user's data from administrators, 
co-tenants, and other attackers is a major challenge. 

To tackle this problem, research has been done to build ``encryption-in-use'' mechanisms that greatly 
improve security by preventing the attackers and even the cloud operators from ever seeing the data in clear.  %There is an inherent security vs performance tradeoff here \cite{GO96,CT14}.
A lot of work has been done on improving the security and performance on a subset of SQL operations as systematized in the survey by \cite{sokencdb}, but only a handful of systems are complete and evaluated at scale.
%Systems built using ``encryption-in-use'' techniques~\cite{cryptdb,mylar,cipherbase,arx} greatly 
%improve security by preventing attackers and even cloud operators from ever seeing the data in clear. 
%For instance, a number of approaches proposed versions of encrypted database systems. 
%\dnote{a sentence to narrow down the encrypted DBs of interest to cryptb and ones henceforth}
 %The core idea behind these systems is to encrypt the data and perform processing ``over the encrypted data''. 
 The state of art encryption-in-use database systems which have been evaluated at scale can be divided into two main categories: 
\begin{enumerate}[label={(\Alph*)}]
\item systems built using advanced encryption schemes that allow to perform operations over the ciphertexts~\cite{cryptdb,arx,blindseer}, and
\item systems that leverage a trusted processing device (e.g., FPGA, IBM secure co-processor) to perform operations~\cite{cipherbase,trusteddb,oblidb}. 
\end{enumerate}
A practical encrypted database design is evaluated in terms of the following four aspects: 
\begin{itemize}
\item security: leakage profile and security assumptions. Leakage profile characterizes the amount of data leakage introduced by the design. Security assumptions include the mathematical assumptions for the cryptography and the trusted computing base (TCB) and other trust assumptions for the trusted hardware.
\item functionality: the SQL operations and DBMS functions supported.
\item performance: throughput, latency and scalability to large datasets.
\item intrusiveness level: amount of changes to the underlying DBMS. %This is tied to the DBMS functions supported, with a
\end{itemize} %The need for the first three aspects do not require motivation,
%It is extremely challenging to achieve a good rating in all even when focusing only on the first three aspects. Further, the adoption of an encrypted database design relies on it building on top of a practical DBMS without making extensive changes to the core DBMS components, and hence non-intrusiveness also plays a big role in the practicality of a design.
%\dnote{specify that we compare only with complete databases and a lot of work is done to support parts, and refer to the SoK for those references.}
%For instance, to support encrypted search without any leakage, there is an inherent logarithmic overhead in performance \cite{GO96, CT14}. 
%So, all these systems %in the category (A) had to balance between large leakage profiles, support for limited query functionality, and large performance overheads. 
%Among the available designs, all 
CryptDB \cite{cryptdb} is a seminal work in this area using property-preserving encryption schemes to execute queries over encrypted data. But, these schemes do not offer strong security and when used in multiple columns they are found to leak extensive information for real-world datasets \cite{inference,breakingmylar}. Also, \cite{cryptdb} requires extensive computations (re-encryption of entire columns) on a trusted proxy or the client to support all the SQL queries. The other systems using advanced encryption schemes either have a very limited functionality \cite{blindseer,ospir1,sisospir} or incur heavy computational and storage overheads \cite{arx}. 

Cipherbase \cite{cipherbase} offers a scalable design for transactional workloads with a strong leakage profile and complete SQL support, by leveraging on trusted hardware. But, the system uses FPGAs as its trusted hardware and hence has the following security implications: (i) an initial trusted and on-premise key loading phase is required for every FPGA device used,
(ii) a huge trust is placed on the FPGA``shell'' layer \cite{awsf1shell} implemented by the cloud operators which monitors the user operations on the FPGA to ensure the safety of the device. As such, significant research is required to use FPGAs as \textit{trusted} hardware in cloud-based applications. The other trusted hardware based systems \cite{trusteddb, oblidb} offer improved leakage profile but only at the cost of extensive DBMS changes, much larger TCB and huge performance overheads for large transactional workloads. %, and it is important to study  alternative solutions. 

In this work, we study how to build an encrypted database system from a standard CPU leveraging the Intel Software Guard Extensions (SGX) instruction set~\cite{sgx}. 
SGX enables the creation of a small encrypted memory container (enclave) that can be accessed only by a predefined 
trusted code. The content of the enclave is protected from 
untrusted applications and even the system administrators, OS and hypervisor. 
Also, SGX is available in all the recent and future releases of Intel CPUs.
Hence, SGX offers a great direction for protecting applications in cloud environments.
% Unfortunately, it is not immediately clear how to design an encrypted database on top of Intel SGX. 
But, SGX has its own set of restrictions. %, which makes it unclear whether any of the previous designs based on trusted hardware would be applicable to SGX.
It requires rewriting of applications by partitioning code into trusted and untrusted segments.
Also, there is a $90$ MB bound on ``secure'' memory to run the trusted enclave code, which is not nearly enough for 
even medium size database workloads.\footnote{Although various SGX extensions are promised by Intel in future releases with larger secure memory, they are not available in the market yet and unclear when they will be. We also argue in the paper that these extensions should not affect our conclusions on the architecture of an encrypted DBMS with SGX.} 
%Databases need to scale to support arbitrary querying of gigabytes, or terabytes of data. 
Additionally, SGX is vulnerable to memory, cache and other side-channel leakages, 
lacks syscalls and IO support, and incurs high overheads for switching between enclave and non-enclave modes, which further limit the complexity and functionality of the trusted enclave code.
As such, one cannot take a DBMS system and naively try to ``run it in an enclave''. 
%Another challenge when building a database with SGX is to make sure that
%compromises in DBMS codebase or its authentication mechanisms do not result in 
%data leakage. %\dnote{Is this sentence to motivate our use of authentication enclave? Seems a bit imprecise. We do not do anything to prevent this.}
But, it is important for an encrypted database design to get around these limitations without having to 
make extensive changes to the underlying DBMS, while still achieving the performance, security and functionality goals. 
Also, it is not clear whether a design that works well for another trusted hardware can be ported to SGX while preserving the end-to-end security guarantees, since each hardware has its unique set of security and usability requirements. % \dnote{multi-tenant}
\subsection{Our contributions}
\begin{figure}[h!]
\centering
\includegraphics[scale=0.75]{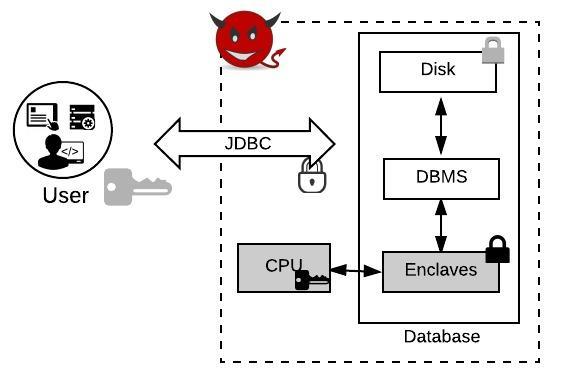}
\caption{High-level architecture overview of StealthDB}
\label{fig:overview}
\end{figure}

%In our quest towards designing an encrypted database system leveraging Intel SGX, 
%We first identify a desired list of security and functionality properties, 
%along with a set of design goals and constraints (Section~\ref{sec:wishlist}). 
%The main security goal is that we want no information leaked i.e., provide semantic security to all the data that is not used during the execution of a query, 
%even to an adversary having persistent access to the memory.
%Also, as we mentioned, we permit minor or no changes to the underlying DBMS.
%We refer the reader to Section~\ref{sec:wishlist} for a detailed list of our requirements and design constraints. 
\vspace{-10pt}
\paragraph*{Design choices with SGX} We first investigate three possible design choices for an encrypted database with SGX in Section~\ref{sec:designs} by varying the DBMS components run inside an enclave.
%The first one runs the whole DBMS inside an enclave using applications that enable running unmodified executables inside enclaves. The second one runs the query execution logic inside an enclave in a similar manner and the rest of the DBMS outside in the untrusted zone. The third one is inspired by \cite{cipherbase} and performs only the \textit{primitive} operations inside an enclave.
Through a set of benchmarking experiments, we identify a design that works best for our design goals (Section~\ref{sec:architecture}). We develop on that to get the \name{} design.  %The first two choices will incur high performance overheads. Even their security is not clear since
\paragraph*{StealthDB} The \name{} system provides a complete SQL support, strong end-to-end security guarantees and performance with minimal changes to the underlying DBMS. A high-level overview of our system is presented in Figure~\ref{fig:overview}. 
\name{} uses AES-CTR, a semantically secure encryption scheme to encrypt all the data items in the database.
During query execution, the client encrypts the query string and sends the ciphertext to the server. We implement a query parser inside an enclave, which first decrypts the ciphertext to get the query and parses the query to output a version with all the constants encrypted. For example, when a client sends \textsf{ENC(select * from item where name = `John')}, it is converted to \textsf{select * from item where name = `ENC(John)'} by our enclave parser. To support queries of this form, we define encrypted datatypes and implement the operators over these datatypes inside an enclave. We make the operators data-oblivious \cite{OSF+16} to protect against SGX side-channel attacks. We also encrypt the index file pages before they are written to disk. These changes are not intrusive and hence enable \name{} inherit the functionality of the underlying DBMS completely.
\paragraph*{Security} \name{} offers a stronger leakage profile compared to the prior complete encrypted database systems. A snapshot adversary \cite{RY10,GR05,BNPSS11,BZBKL12,verizonbreach} learns only the ``shape'' of the database which includes the dimensions of the data structures maintained by the DBMS, along the recently collected query log information. An adversary with persistent access to memory and disk learns the inequalities (<, >, =) between the encrypted values in the indexes which are compared during the query execution,  along with the query access pattern which includes the position of the result records in the database. 
%For example, when executing the query \textsf{select * from item where name < `ENC(John)'}, a persistent adversary learns at most the following: the inequalities between \textsf{name}s in \textsf{item} which are lexicographically before John and the position of these records in the database. We do not reveal anything more about the data items in the database.
In general, the enclave code can be thought of as providing a black-box access to the DBMS to perform the computations on encrypted data values and obtain the output (encrypted or unencrypted depending on the specification), without leaking any other information about the input data values. We explain our leakage profile in more detail in Section~\ref{sec:security}, and this profile matches the state-of-art (the strongest version in \cite{cipherbase}) when providing either reasonable performance\footnote{From our experience talking to the industry on the possible adoption of \textsf{StealthDB}, 50\% to $2\times$ overhead in \textit{performance} is a reasonable penalty for the benefit of security against untrusted cloud operators.}
 or intrusiveness levels for large transactional workloads. %A concurrent work of ours \cite{oblidb} builds on SGX to support B+ tree indexes without leakage, but they incur
Also, our TCB just includes the processor, the enclave code along with the SGX hardware and the attestation procedure. Our clients use the SGX attestation procedure to attest the correctness of the enclave code before issuing queries. This combined with the simplicity of the enclave code reduces the trust to be placed on the enclave code. %It is not clear how to achieve this with an FPGA. Second, we encrypt the index and log files stored on disk since they reveal significant information even to an adversary which gets a snapshot access to the disk \cite{GRS17}.

\paragraph*{Evaluation} We implement our design on top of an existing Postgres DBMS. %Our source code is available at \url{https://github.com/cryptograph/stealthdb}.
 %via a series of extensions and add-on modules. 
Our new encrypted datatypes and the corresponding UDFs are added as extensions in Postegres \cite{pgext}. The only component that needs modifying the Postgres code is to encrypt/decrypt the index files when they are stored to/accessed from disk and this just needs a three lines change in the Postgres codebase. None of these changes are intrusive, or specific to Postgres.
Hence, this design principle lets \name{} benefit directly from any performance or feature improvements to the underlying DBMS engine. 
Performance-wise, \name{} scales to large datasets with a similar complexity
to an unmodified DBMS engine working on unencrypted data, adding only a tiny overhead for each query. 
Our evaluation results in Section~\ref{sec:implementation} show that the system can process OLTP queries with a 30\% reduction in throughput and $\approx 1$ ms overhead in latency over an unencrypted DBMS
with $>10$M total rows (or 2 GB plaintext) of a TPC-C warehouse database for scale factor $W=16$. 

%\paragraph*{Comparison with other trusted hardware solutions} 
%TrustedDB \cite{trusteddb} has the full DBMS in its TCB. A concurrent work ObliDB \cite{oblidb} builds on SGX to support B+ tree indexes without leakage, but does not scale well to large transactional workloads. %\dnote{oblidb doesnt provide any tpc-c like evaluation, so can I make this claim?}
%The Cipherbase design does not inherently need an FPGA, but every trusted hardware has its own set of security and usability requirements. Hence, it is not clear whether an FPGA based design can be ported to another hardware while preserving the end-to-end security guarantees. % \dnote{multi-tenant}

%Hence, \name{} provides the security guarantee that even an adversary with persistent access to memory learns no information about the data in all the non-indexed columns. We do not modify the index building and usage procedures of the underlying DBMS, and hence the structure of the index tree in memory reveals some inequalities among the elements in the tree, even though the individual elements remain encrypted with AES. Sections~\ref{sec:wishlist} and \ref{sec:security} present more details of our leakage profile.
%We call the resulting system StealthDB -- a database engine where queries are processed in stealth mode. 
\dnote{multi-client}
\dnote{talk a bit more about the parser later}

\section{Background on Intel SGX}
%\label{sec:prelims}
\label{sec:sgx_background}

In this section we give a brief introduction to Intel Software Guard Extensions (SGX).
We refer the reader to~\cite{sgx,CD16} for more details on SGX.
Intel SGX is a set of new x86 instructions that enable code isolation within virtual containers called enclaves.
In the SGX architecture, developers are responsible for partitioning the application into enclave code
and untrusted code, and to define an appropriate I/O communications interface between them. 
In SGX, security is bootstrapped from an underlying trusted processor, but not trust in a remote software stack.
%On the high level, to a user the SGX hardware presents the $\Load(\program)$ and $\Execute(\Enclave_\program,\inp)$ functionalities.
On the high level, the SGX hardware presents the following two functionalities to a user:
\begin{itemize}
\item {$\Load(\program)\rightarrow(\Enclave_\program, \phi)$}: creates an enclave with an identifier $\Enclave_\program$ and loads the program $\program$ into it. It then produces a proof $\phi$ that the intended program $\program$ (and initial data) has been loaded into the enclave.
\item {$\Execute(\Enclave_\program,\inp) \rightarrow (\out,\psi)$}: given an enclave handle $\Enclave_\program$ (corresponding to an enclave with a program $\program$), $\Execute$ runs it on an input $\inp$ and produces a tuple constituting of the output $\out$ and a proof $\psi$. A client can use $\psi$ to verify that $\out$ was produced by the enclave $\Enclave_\program$ executing with $\inp$.
\end{itemize}
%The load function creates an enclave with an identifier $\Enclave_\program$ and loads the program $\program$ into it. A client receives a proof $\phi$ that its intended program $\program$ (and initial data) has been loaded into an enclave.
%The proof $\phi$ can be used by the client to attest that the right program 
%has been loaded inside an enclave with respect to a measurement (hash) of that program. 

% via an attestation procedure.
%Code loaded into enclaves is measured by SGX during initialization (using SHA-256) and signed with respect to public parameters.
%The client can verify the measurement/signature pair to attest that the intended program was loaded via the Intel Attestation Service. 
%We note that the underlying schema, Intel Enhanced Privacy ID (EPID)\cite{intelepid} is a group signature scheme, that allows the platform to produce signatures that are anonymous and unlinkable.\cnote{can we use the group sig scheme somehow?  if not, probably not worth mentioning.}

%\emph{$\Execute(\Enclave_\program,\inp) \rightarrow (\out,\psi)$}.
%The execute function is given an enclave $\Enclave_\program$ handle (corresponding to an enclave with a program $\program$), it then runs it on an input $\inp$, to produce a tuple constituting of the output $\out$ and a proof $\psi$ which the client can use to verify that the output $\out$ was produced by the enclave $\Enclave_\program$ executing with $\inp$.

There are three main functionalities that enclaves achieve: isolation, sealing and attestation. We provide a high-level description here. Please refer to \cite{CD16,iron} for more detailed and formal descriptions.
%\begin{enumerate}

{\textit{Isolation}}: code and data inside the enclave protected memory cannot be read/modified by any process external to the enclave. 
%SGX does this by isolating enclave code and data in the Processor Reserved Memory (PRM), referred to as Enclave Page Cache (EPC), which is a subset of DRAM that gets set aside securely at boot time. Cache lines read into the processor cache from the EPC are isolated from non-enclave read/writes via hardware paging mechanisms, and encrypted/integrity checked at the processor boundary. Cryptographic keys for these operations are owned by the trusted processor. Thus, data in the EPC is protected (privacy and integrity-wise) against certain physical attacks (e.g., bus snooping), the operating system (direct inspection of pages, DMA), and the hypervisor. 

{\textit{Sealing}}: data passed to the host environment is encrypted and authenticated with a \textit{Seal Key} that is specific to the enclave identity and derived from a hardware-resident \textit{Root Seal Key}. 
%Every SGX processor has a \textit{Root Seal Key} that is embedded during the manufacturing process. 
%An enclave can %use the \texttt{EGETKEY} instruction to 
%derive a \textit{Seal Key} that is specific to the enclave identity from the Root Seal Key and this Seal Key can be used to encrypt/authenticate data and store it in untrusted memory.
%Sealed data can be recovered by the same enclave even after enclave is destroyed and restarted on the same platform. 
%But the Seal key cannot be derived by a different enclave on the same platform or any enclave on a different platform.
%We will use the following $\Seal$ and $\Unseal$ algorithms:
%\begin{align*}
%\Seal(\aad, \msg) &\rightarrow \sct\\
%\Unseal(\aad, \sct) &\rightarrow \msg / \perp
%\end{align*}
SGX uses AES-GCM to encrypt $\msg$ using the Seal key of the enclave calling the function. %Here, $\aad$ is the \textit{additional authentication data} which is included as a part of the MAC step to provide integrity but not encrypted along with $\msg$. We will ignore the $\aad$ argument when there is none.

{\textit{Attestation}}: a special signing key and instructions are used to provide an unforgeable report attesting to code, static data, and (hardware-specific) metadata of an enclave, as well as outputs of computations performed inside the enclave. There are two forms of attestation: \emph{local} and \emph{remote}.
\begin{itemize}[leftmargin=*]
\item \textit{Local attestation.} An enclave $A$ uses local attestation procedure to generate a \emph{report} and attest to another enclave $B$ on the same platform. 
%Since enclaves on the same machine share the same Root Seal Key, the enclave $A$ uses a special instruction which creates a MAC of its measurement and its metadata (along with additional optional data provided as input to the instruction) with a \textit{Report Key} corresponding to the enclave $B$ derived from the Root Seal Key. The resulting MAC is called a \emph{report}. Now, the enclave $B$ can verify the report by deriving the same Report Key from the Root Seal Key.

\item \textit{Remote attestation.} Remote attestation procedure generates a report specific to an enclave called \emph{quote} that can be verified by any remote party. 
%Roughly, an enclave first local attests to a special enclave called the Quoting Enclave (QE), sending it a report. The QE verifies local reports and if valid, signs the same underlying data with a private key for an anonymous group signature scheme called Intel Enhanced Privacy ID (EPID) \cite{EPIDKeyProvisioning}. The QE obtains this private key during through a protocol with the Intel Provisioning Server upon device initialization.  The resulting signature is called a \textit{quote}. 
%This protocol includes a symmetric authentication involving Root Provisioning Key that was embedded in the device during the manufacturing process and also shared with the Intel Provisioning Server. 
%Currently, the remote party requires contacting the Intel Attestation Server to verify quotes, though in principle this could be done by any verifier that has the group public key.
\end{itemize} 
%\end{enumerate}

\emph{Key establishment during attestation.}
Key establishment between two enclaves or between an enclave and a remote party can be accomplished on top of the local/remote attestation process. An enclave can send the key shares (for eg., a Diffie-Hellman key share $g^a$) and include them as the \textit{additional authentication data} to MAC. Thus attestation provides authenticity and integrity to the key share from the enclave. In our system, we will very often run the key establishment phase on top of local/remote attestation to establish a secure channel for communication between two enclaves or between an enclave and a remote party using the established shared secret key. %We will use the following two pairs of function calls to achieve these tasks:
%\begin{align*}
%\LAKE_\src(\destenclave) &\rightarrow k / \perp\\
%\LAKE_\dest(\srcenclave) &\rightarrow k / \perp\\
%\RAKE_\src(\destenclave) &\rightarrow k / \perp\\
%\RAKE_\dest(\src) &\rightarrow k / \perp
%\end{align*}
%Here, $k$ is the key established between the source and destination enclaves if the attestation completes successfully, and $k$ will be used to encrypt the further communication between them. (During remote attestation, the $\src$ need not be an enclave).

\emph{SGX TCB.} 
SGX stands out in that its TCB consists only of the CPU microcode and privileged containers, however it also requires the user to trust in Intel's key management infrastructure for signing microcode and various service enclaves. In particular, we must trust that the root seal keys embedded into devices are not leaked from the manufacturing facility, and that the Intel Provisioning Server safely manages root provisioning keys as well as other master secret keys.%%Note: 'EPID' changed to 'other'

Although SGX prevents an adversary from directly inspecting/tampering with the contents of the EPC, it does not protect against multiple software-based side channels. 
%As mentioned earlier, SGX enclaves share hardware resources with untrusted applications and delegate EPC paging to the OS.
Correspondingly, the literature has demonstrated attacks that extract sensitive data through hardware resource pressure (e.g., cache~\cite{BMD+17,SWG+17,foreshadow,cachequote}, thread scheduling~\cite{WKPK16} and branch predictor~\cite{LSG+17}) and the application's page-level access pattern~\cite{XCP15}.
Many of these works also provide fixes for their attacks with varying overheads and need to be patched by Intel. For the application's page-level access pattern though, it is  up to the application developer to design data-independent memory accesses for the data to be secure.
%Second, since EPC paging is still managed by the OS, which pages are swapped and when (the page-level memory access pattern) is revealed to the OS. 
%These data dependent page access pattern leakages in conjunction with an offline code analysis can be used by an adversary to infer sensitive enclave data.
%The simplicity of our enclave code also enables various formal verification techniques.
%\emph{EPC scope.} 
%Since the integrity verification tree for EPC pages is located in the EPC itself (in VA pages), SGX does not support integrity (with freshness) guarantees in the event of a system shutdown~\cite{rote}.
%More generally, SGX provides no privacy/integrity guarantees for any memory beyond the EPC (e.g., non-volitile disk).
%Ensuring persistent integrity for data and privacy/integrity for non-volitile data is delegated to the user/application level. 

\section{Platform Overview}\label{sec:platform}

%This section describes our setting, platform details and security model which we study.
\subsection{Usage Model.}

We work with the following setting.
A data owner aims to store and process data securely on a remote untrusted SQL database server.
She authorizes clients by issuing them \textit{credentials}, and wants to support the authorized clients to issue queries to the server.
The server maintains a \textit{credential database} for the authorized clients in an encrypted form.
Each client authenticates to the server using its credentials, which will enable the client to issue
its permitted queries to the database.
The server in our model is equipped with a secure processor, such as Intel SGX. 
Hence, the server can be identified with some ``platform-key'' established by Intel SGX. 
%This key will be used during the attestation protocols. 
The data owner and clients engage in the attestation of SGX enclaves in the server and on successful attestations, transfer any secret or sensitive material (master key, credentials, queries, etc.) to those enclaves via secure channels. 

%then engages in a remote-attestation \textit{plus} key-establishment phase with an SGX enclave in the server to set up a secure communication channel using the established shared secret key.
%If it succeeds, he sends the credential database through the secure channel to the enclave.
%The enclave generates and holds long term secret key(s) used to encrypt the data on the database server. 
%Now, to issue queries, client runs a remote-attestation \textit{plus} key-establishment with the enclave and uses the resulting secure channel to send encrypted queries. 
%The server, with the help of the enclave, will be able to efficiently execute the encrypted queries and return encrypted results back to the client.
%The data owner sends an encrypted version of its database to the server.

\subsection{Threat Model}  \label{sec:threatmodel}
%\dnote{try to understand from a new reader's perspective and re-write.}
\name{} provides security against passive adversaries. A passive adversary does not inject malicious code or alter the program execution in any way. But, it can read the contents of the memory, disk and all the communication, and hence may \textit{passively} attempt to learn additional information from the data they observe. 
%Hence, we do not protect against adversaries who can modify the query to be executed after it is received by the DBMS. 

There are two dimensions in which we analyze the threat model for our system. The first dimension is about the extent of access: adversaries restricted to monitoring the disk accesses versus the adversaries monitoring both the memory and disk accesses in the system. 
The second dimension is about the duration: adversaries getting snapshot accesses to memory or disk versus the much stronger ones which get persistent access to memory and disk. A snapshot attack might be due to a memory dump or some cold-boot attack by a malicious cloud provider or by a co-located client running on the same cloud server as the victim process which gets occasional access to the memory of the entire system due to access control bugs. SQL injection attacks \cite{mysqlexploit,osexploit,verizonbreach}, VM attack leaks \cite{RY10,GR05,BNPSS11,BZBKL12}, disk theft and a ``smash-and-grab'' after a full system compromise \cite{verizonbreach} are some real-world examples of snapshot attacks \cite{GRS17}. 
\section{Designing an Encrypted DB}

In this section, we describe a few design goals we set out to achieve for our system. Then, we discuss and experiment with a few 
possible design choices possible when building an encrypted database from SGX. 

\subsection{Design Goals} \label{sec:wishlist}
%There is usually a three way trade-off between security, functionality and performance while designing an encrypted database. We also stress the need for the fourth dimension of non-intrusiveness which precludes significant changes to the underlying DBMS.
%Often, an optimal trade-off depends on the underlying data and query types that the DBMS needs to support. 
The focus of \name{} is on building a scalable encrypted database system that can support arbitrary query types, with a reasonable leakage.
Construction of an encrypted DBMS with a complete SQL support under any meaningful notion of security is an uphill task in this world where the proposed attacks \cite{breakingmylar,GRS17,KKNO16,GLMP18} completely dismantle the security of even the constructions with limited functionality (like searchable encryption) which had, what was thought to be, minimal leakage (reveal just the locations of the results of each query). There has been extensive research to secure subsets of SQL operations \cite{sokencdb}, but a proposal can be included in a real world DBMS only if it is compatible with or provides a complete support of the DBMS tasks. For instance, the CryptDB design was part of or inspired many real-world systems \cite{cryptdb,msalwaysenc,seeed,encbigquery} due to an almost complete DBMS support. In this regard, we set our design goals as follows: \dnote{what is the purpose of the goals section? should we define only those goals that we achieve, or can we state ideal goals and tell where we miss?}
\begin{itemize}
\item Functionality goal: complete support to the SQL functionality of the underlying DBMS.
\item Non-intrusiveness goal: minor modifications to the core DBMS operations of the underlying DBMS, for the encrypted database to retain the DBMS properties. If the underlying DBMS is ACID compliant, supports triggers and stored procedures, so should the encrypted database.
\item Performance goal: high throughput and low latency when scaling to large datasets.
%encrypted DBMS query runtime that scales identically to the underlying DBMS engines. That is, any query that takes $T$ time to execute in a native DBMS, should take at most $c \cdot T$ time to execute over encrypted data, for some constant $c$. \dnote{say ``high throughput and low latency'', or even just ``high throughput'' since TPC-H numbers are bad anyway.}
\item Security goal: We will start by stating the security goals informally:
%\dnote{option 1}
%\begin{itemize}
%\item for both snapshot and persistent adversaries monitoring the disk accesses, provide semantic security to all the data and indexes
%\item for snapshot adversaries on memory, provide semantic security to all the in-memory data
%\item for persistent adversaries on memory, provide semantic security to the data not used by the queries that were executed during the monitoring by the adversary. For the data used by the queries, reveal only the minimal information about the data required for query execution.
%%semantic security of cold in-memory cached data (i.e. when queries are not executed)
%\end{itemize}
%\dnote{option 2}
\begin{itemize}
\item a snapshot adversary on both memory and disk should learn no information about the individual data items.
%\item a persistent adversary on both memory and disk learns no information about data that are not part of the query results, other than that they are not part of the query results. Even for the data part of the query results, the leakage should match or be stronger than the previous works supporting complete SQL.
\item a persistent adversary on both memory and disk learns no information about the encrypted data that are not compared when the queries are processed, other than that they are not part of the query processing. Even for the data of the query execution, the leakage should match or be stronger than the previous works supporting complete SQL.
%semantic security of cold in-memory cached data (i.e. when queries are not executed)
\end{itemize}
We will later study the security for each proposed design. And, the leakage profile of the chosen \name{} design will be detailed in Section~\ref{sec:security}.
\end{itemize}
There is an inherent trade-off here between security and performance which will influence our design choices.
There is a lower bound of logarithmic overhead in performance \cite{GO96, CT14}, just to support encrypted search without any leakage. This also translates to the trade-off between efficiency and the information leakage during the index building and usage. Moreover, we also aim to design secure versions of arithmetic and other operators to support SQL completely. Hence in this work, we lean towards achieving a good performance for large transactional workloads, while trying to achieve the best security possible for that performance.
% We refer the reader to Section~\ref{sec:security} for more details on the leakage profile of \textsf{StealthDB}.
%\snote{can we quantify what the constant c is?}
%\dnote{give a summary here of what existing systems can/cannot do with respect to these goals.. possibly a table.}
\begin{figure*}[t]
\centering
\includegraphics[width=17cm,height=12cm,keepaspectratio]{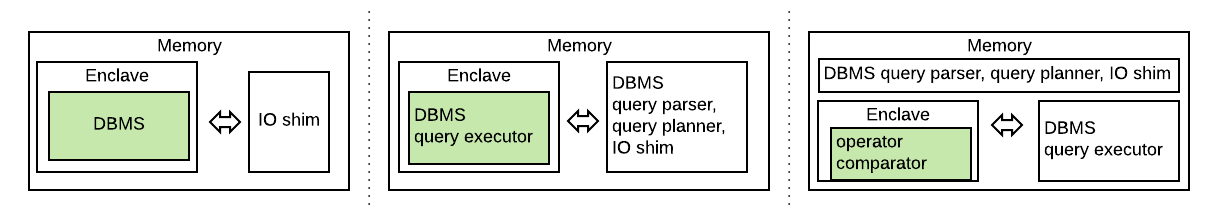}
\caption{Three alternative design choices for an encrypted database with SGX.}
\label{fig:designs}
\end{figure*}

\subsection{Designing an Encrypted DB from SGX}
\label{sec:designs}
We consider three design choices and evaluate them on a few micro experiments to help us understand how to build an encrypted 
database system with SGX. The design choices are summarized in Figure~\ref{fig:designs}.
We envision that in all three design choices data is encrypted on disk using a semantically secure encryption scheme. 
The designs differ in how queries are executed over the data. 

The first, most obvious design would be to run the entire DBMS inside an enclave (left figure in~\ref{fig:designs}). 
The data would be read from disk, decrypted transparently and then the DBMS would perform all necessary operations inside an enclave. 
However, SGX is not well suited for this task for a few reasons that we outlined earlier. The first issue is that SGX does not support IO or syscalls, 
so an additional outside shim layer would need to be exposed to talk to the kernel level, and the application 
dependencies need to be loaded inside (or outside via shim) an enclave. 
It is \textit{feasible} to get around this issue using recent works such as Haven~\cite{haven}, Scone~\cite{scone} and Graphene~\cite{graphene,graphenesgx}.
They initiate the research in loading unmodified executables into enclaves. The second issue is that SGX is currently 
limited to 90 MB of working memory and significant penalties appear when going beyond that limit~\cite{eleos}.
Future releases of SGX promise larger enclave sizes. However, the Merkle tree integrity protection 
for each memory page to prevent replay attacks does not scale well to larger enclaves.
These two issues would result in heavy performance overheads on transactional workloads for this design.\footnote{We do not do a direct performance evaluation for this design, but the design that we discuss next which runs much less operations inside an enclave already has high overheads.}
But, this design can have better confidentiality guarantees when the SGX-based side-channels are addressed.
%If these two issues were resolved, this design offers no information leakage to a snapshot attacker, since the memory used by the DBMS inside an enclave always remains encrypted.
%But, this design would keep a very large TCB inside the enclave: the entire DBMS engine, any communication
%logic with the ``outside world'' and dependencies. 
%Since SGX is vulnerable to numerous side-channels, very custom modifications to the DBMS are
%needed to prevent these attacks and make the code \emph{oblivious}. Hence, this design is not very promising.
%\dnote{no leakage to snapshot, since indexes are encrypted; but a persistent adv can learn access patterns.... }

\begin{figure}[h!]
 \centering
  \includegraphics[width=0.4\textwidth, scale=0.9]{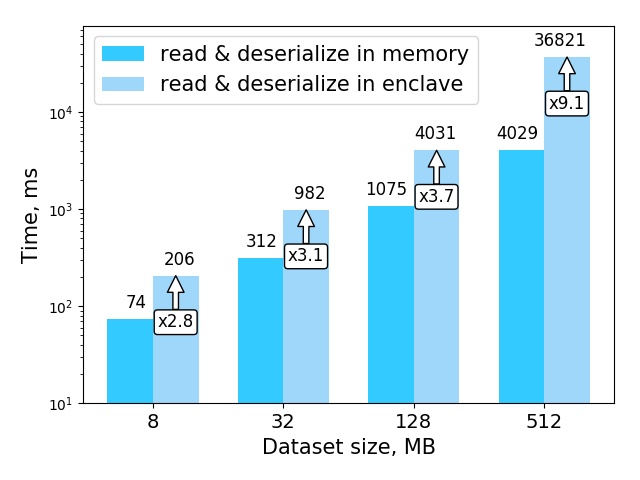}
 \caption{Initialization time comparing in memory and in enclave deserialization for different dataset sizes.}
\label{fig:initTable} 
\end{figure}

The second design we consider (middle figure in~\ref{fig:designs}) keeps most of the DBMS in the untrusted
zone. However, it places the query execution logic in the enclave. That is, when a query needs to be executed,
individual tables can be brought in to the enclave to perform selections, projections, joins, etc. 
The query plan, I/O and other DBMS parts remain in the untrusted memory. 
In terms of scalability, this design suffers from the same problems as the previous choice due to limited secure memory. 
Also, tables and indexes need to be read from disk, deserialized and then loaded into enclave. In Figure~\ref{fig:initTable}
we show that the performance overhead for performing just this step (read and deserialize) inside an enclave is around $3\times$ when the dataset fits 
within an enclave, and goes up to $9\times$ for large datasets. In terms of security, the query processing logic 
would still need to do the non-trivial task of addressing the SGX side-channels. Finally, partitioning a DBMS to support this architecture is also a challenging task. 
%\dnote{what abt leakage to snapshot?
%same argument as above for persistent..
%for an ideal sgx, both these two also provide integrity, but..}

%\begin{figure}[h!]
% \centering
%  \includegraphics[width=0.5\textwidth, scale=1.2]{table1.jpeg}
% \caption{Initialization time comparing in memory and in enclave approaches for different datasets}
%\label{initTable} 
%\end{figure}

\begin{figure}[h]
 \centering
  \includegraphics[width=0.4\textwidth]{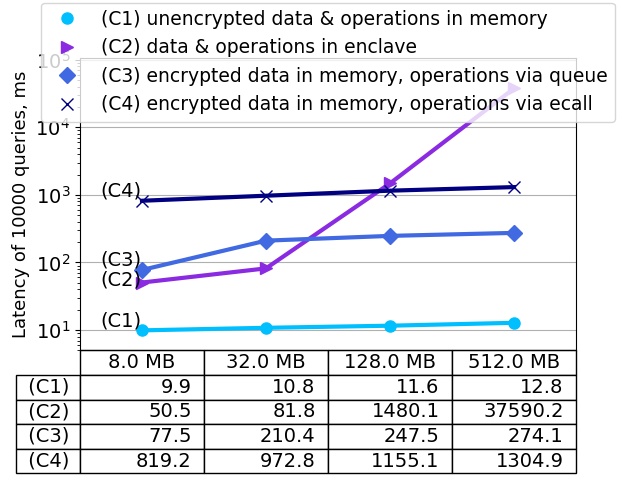}
 \caption{Latency to execute random binary tree searches comparing different approaches. Two different implementations of the partial approach: comparison function as trusted ecalls and the exit-less communication via a queue for transferring data to/from an enclave.
 %{\color{red} SNote: swap order to rows. c1, c2, c3, c4.. also change c1: unencrypted data and operations ..., c2, c3: encrypted data in ...,}
}
\label{fig:initLatTable} 
\end{figure}

In the third design, we keep most of the DBMS in the untrusted zone, and the dataset would reside in the untrusted memory with the data items encrypted individually.
At the lowest level of the parsed query tree, each query is eventually broken down into some \textit{primitive} operators (e.g., $<=, >=, +, *$) 
over individual data values. To perform operations over encrypted data in this design, we transfer individual data item(s) 
to an enclave, followed by the decryption of input, the operator function and the encryption of output inside the enclave. 
%Hence, the data relation structure produced by the query is kept in the untrusted memory. 
The advantage of this design is that the communication with the disk and network layers would remain unchanged.
Overall, minimal changes to the DBMS are needed -- one only needs to change how primitive operators on 
data values are performed. 
Also, the amount of code/data inside an enclave will remain a very small constant.
This keeps the TCB very small, and it is easy to make it data-oblivious. Hence, we build on this design idea in Section~\ref{sec:architecture}.
However, this design leaks relationship between encrypted data values during query execution in this design as discussed in Section~\ref{sec:security}. 

%In the last design, we keep most of the DBMS in the untrusted zone.
%Communication with the disk and network layers would not change in this design.
%However, to perform operations over encrypted data, individual data item(s) would be 
%transferred to an enclave, decrypted, followed by operator function and encryption. 
%That is, at the lowest level, each query is eventually broken down into operators (e.g., $<=, >=, +, *$) 
%over individual data values. Hence, we may only execute these operators in an enclave
%and keep data relation structure produced by the query in the untrusted memory. 
%For instance, suppose we need to execute a query that selects values $X$ from a column such that $X < 100$. 
%Values in the query would first be encrypted and the query would ask to select values from a column such that $E_i < E^*$, where $E^*$ 
%is an encryption of $100$. 
%Then for each encrypted value $E_i$ in the column, $E_i$ would be compared to $E^*$ inside an enclave.
%The result of this comparison would be stored in the untrusted memory. 
%In this design, minimal changes to the DBMS are needed -- one only needs to change how primitive operators on 
%data values are performed (see Section~\ref{sec:architecture}). 
%Also, in this design the amount of code/data inside an enclave can remain very small (constant),
%since the entire datasets would reside in the untrusted memory encrypted. 
%This keeps the TCB very small, and easy to make it data-oblivious. 
%However, one leaks relationship between encrypted data values during query execution in this design. 

In Figure~\ref{fig:initLatTable}, we compare performance of performing B-tree searches 
over database indexes in later two design choices. As expected, one can see
that performing a search when an entire B-tree is loaded inside an enclave does not scale to 
larger datasets. (However, it performs well when the tree size is very small and can be fit entirely into an enclave.) 
In the third design, when the B-tree is kept encrypted in the untrusted memory but individual comparisons
are executed in an enclave, we see up to $100\times$ overheads compared to performing the search over unencrypted data.
This can be explained by high switching costs for ecall/ocall functions, which are used for enclave entry/exit. 
Using an \textit{exit-less} communication mechanism via a shared queue~\cite{eleos}, we can reduce this overhead 
by $5\times - 10\times$.

\section{Architecture} \label{sec:architecture}
\begin{figure*}[t]
\centering
 \hspace*{-0.45cm}\includegraphics[width=16cm,height=8cm,keepaspectratio]{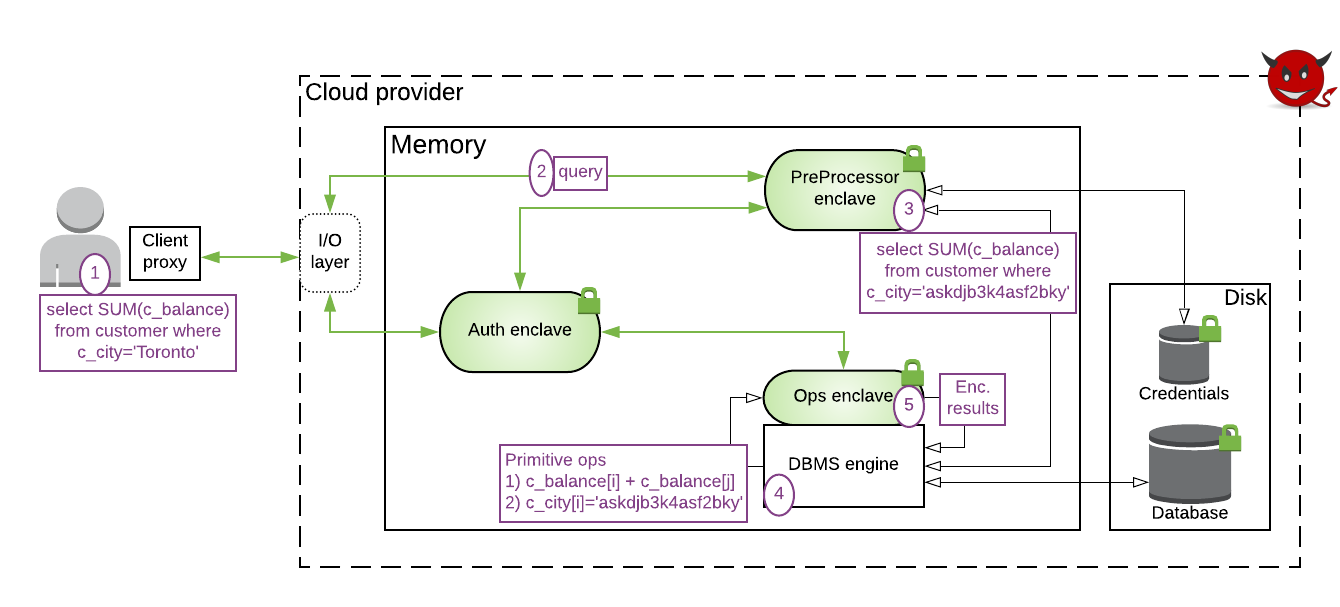}
\caption{\name{} architecture. The life cycle of a query initiating from a client can be traced from steps 1 to 5. The lines with shaded arrows represent encrypted communication between those entities.}
\label{fig:architecture}
\end{figure*}

The architecture of \name{} is presented in Figure \ref{fig:architecture}. As discussed in our third design, \name{} makes minimal changes to the underlying DBMS, with most of our components augmented on top of an unmodified DBMS.
We will now go through the flow of database creation and query life-cycle, and explain each of our components in detail as needed. 

% In particular, \name{} is made up of a relational DBMS \textit{plus} the following:
%\begin{itemize}
%\item a set of guidelines to be used during the database schema design 
%\item augmenting the client authentication mechanism
%\snote{what is ``augmenting''? client proxy? client jdbc driver? give it a name}
%\item augmenting a preliminary query parser to identify and encrypt the data values in the query
%\snote{confusing}
%\item augmenting the set of datatypes and the corresponding set of operators
%\snote{user-defined extension}
%\item a change to the DBMS to encrypt data (index and logs) just before it is written to disk and to decrypt the ciphertext right after it is read from disk.
%\end{itemize}

\subsection{Database creation}
%The creation of a database involves the following steps. The admin designs the database schema, which defines the structure of the database. He then sets the configuration files and runs the database server. Finally, he creates tables and indexes according to the schema and populates the tables with values, if he already has any.
When a database is created, the {database owner} designs a database schema to define the structure of the database. During the schema creation, \name{} allows the owner to identify the columns of the tables in the database which have sensitive information and use our \textit{encrypted datatypes} for those columns.
An encrypted datatype is used to represent values which are the encrypted versions of its corresponding plaintext datatype. For instance, encrypted integers are represented by the encrypted datatype \textit{enc\_int4}. 
\begin{figure}[H]
\centering
\includegraphics[scale=0.4]{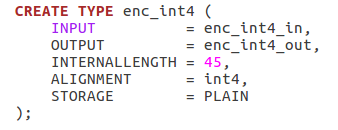}
\caption{Definition of \textit{enc\_int4}}
\end{figure}
%This way, the administrator can also choose the optimal security vs performance tradeoff when using \name{}.
And, a database owner can issue the following command to create a table \textit{item} with two columns of types encrypted integers and encrypted strings:
\begin{figure}[H]
\centering
\includegraphics[scale=0.4]{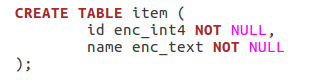}
\end{figure}
%Also, \name{} introduces encrypted operators which perform operations over the encrypted datatypes.

\name{} will encrypt the data values in an encrypted datatype using AES-CTR which is an encryption scheme providing confidentiality of the data values. We will discuss about the key(s) used by this encryption during the DBMS initialization.
%StealthDB supports all basic data-types of a database system outlined in Table~\ref{tab:encdatatypes}. 
%\begin{table}[h]
%\centering
% \begin{tabular}{| c |} 
%\hline\\
%encint\\
%encstring\\
%encfloat\\
%??\\
%\hline
% \end{tabular}
% \caption{List of supported encrypted datatypes}
%\label{tab:encdatatypes}
%\end{table} 
%\dnote{Discuss with Sergey about this table}
 
\subsection{DBMS Initialization}
When the DBMS is started, the following additional steps are performed for StealthDB.

%A data owner who will store his data in the database samples a (master) key at random from the key space. This master key will be used to encrypt their data values in the database. (We do this for simplicity and we will later explain how our design supports an integration with a key management service to enable the usage of different keys for different clients or for different columns in the database.)
%
%The data owner then authenticates the secure enclave running in the DBMS server, and transfers the master key to the enclave if the authentication succeeds. 
%This step is done using the remote attestation service provided by Intel SGX. The data owner program posses public parameters for this authentication and the measurement of the program that the DBMS server has to run inside an enclave. So, when it receives a signed report from the SGX hardware in the DBMS server on the measurement of the program that is run inside an enclave, the client program can check its validity with respect to the expected measurement. 

\paragraph*{Enclaves creation}
\name{} creates three enclaves on the database server: the client authentication enclave $\Auth$, the query pre-processing enclave $\PP$ and the operation enclave $\OP$.
%The full descriptions of the programs that are to be loaded inside these enclaves is described in Figures~\ref{fig:auth},~\ref{fig:pp}, and~\ref{fig:op}, respectively. 
These enclaves are loaded by an \emph{untrusted} DBMS runtime, but our system will later allow to \emph{attest} that the correct code has been loaded into the
enclaves. 
The clients use the remote attestation process and the publicly available measurements (hash) of the enclave code to ensure the correctness of the loaded programs in the enclaves. 
We will defer the explanation of this step and the functionality of these enclaves to the sections below. 

To facilitate the communication between the users and the enclaves, \name{} introduces an I/O layer 
on the server side. Its job is to simply redirect requests between the appropriate enclaves and the DBMS.
%It runs in the untrusted environment.
This will also act as the \textit{wrapper} program for the enclaves helping in processing their I/O requests and system calls.
Note that this layer is outside the SGX TCB, hence it is untrusted and can be controlled by an adversary.

\begin{figure}[t]
\centering
\hspace*{-1.45cm}
 \includegraphics[width=10cm,height=30cm,keepaspectratio]{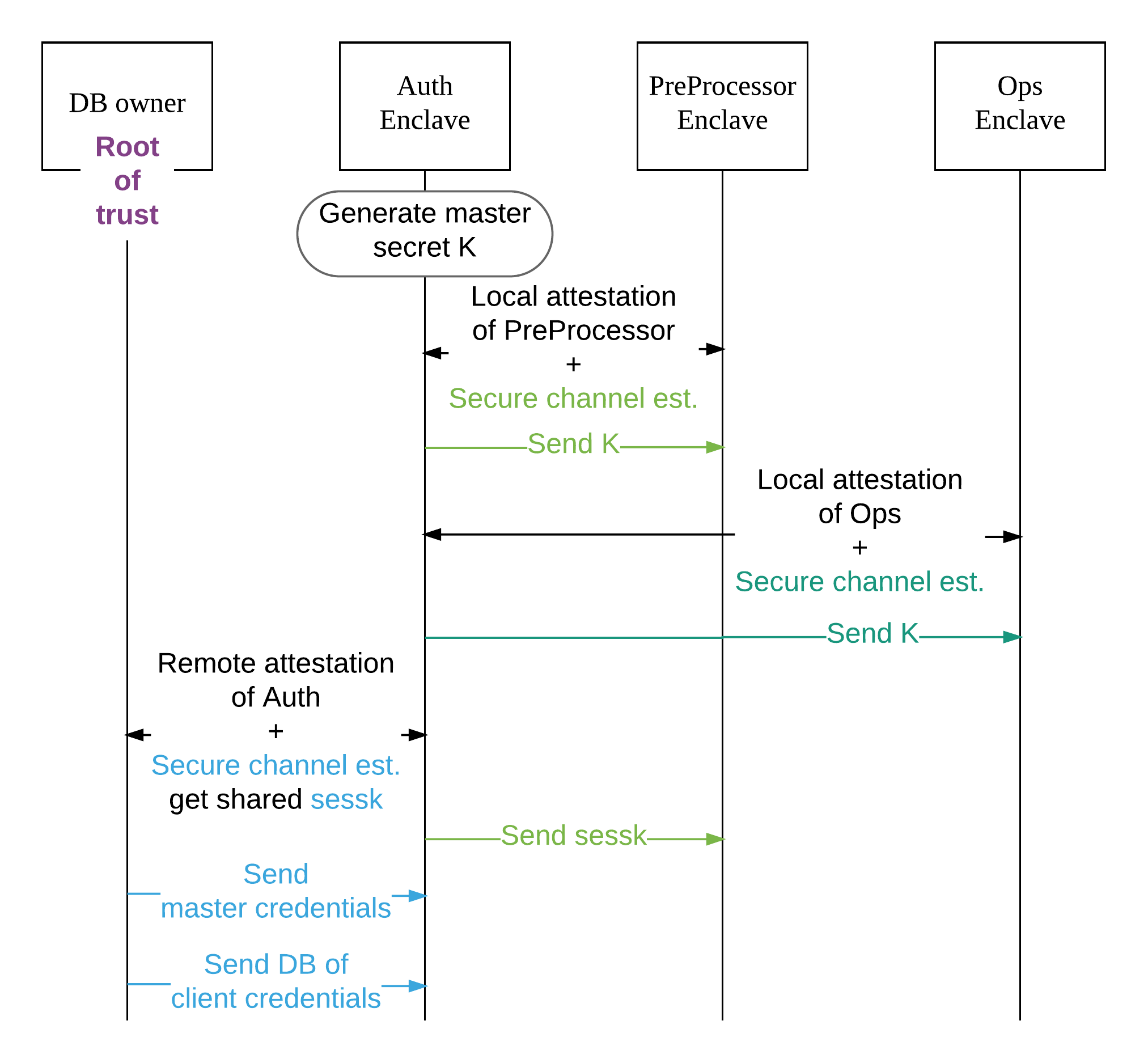}
\caption{The authentication protocol of \name{}}
\label{fig:authprotocol}
\end{figure}

\paragraph*{Key generation}
The  initialization phase also involves generating a master secret key. \name{} performs key generation inside the $\Auth$ enclave. $\Auth$ runs the $\KeyGen()$ function to sample a 128 bit secret key $\key$ at random for the AES encryption/decryption operations. In the current design, this master key $\key$ will be used to encrypt all the data values in the database. We do this for simplicity and our design can be extended to support an integration with a key management service to enable the usage of different keys for different clients or for different columns in the database.

Figure \ref{fig:authprotocol} outlines the key generation and transfer procedures.
The master key $\key$ is then transferred to the $\PP$ and the $\OP$ enclaves as follows. 
When the $\PP$ and $\OP$ enclaves are created, they individually perform a local attestation with $\Auth$ and establish a secure channel with $\Auth$. 
When the attestations succeed and after the secure channels are established, $\Auth$'s $\KeyTransfer()$ function uses the channels to send the master key $\key$ to $\PP$ and $\OP$. (On the other end, $\PP$ and $\OP$ will run their $\KeyReceive()$ functions to complete these steps and receive $\key$). On obtaining $\key$, $\PP$ and $\OP$ use SGX's sealing property to encrypt and store $\key$ for future use.
%\snote{is the name of the enclave ``auth'' or $``auth\_enclave''$. perhaps connect via underscore all enclaves}

\paragraph*{Transfer of credentials}
The final task of the initialization phase involves transferring the client credentials and access policies to $\Auth$. A client (proxy) will authenticate to $\Auth$. And, from the point of view of the DBMS, $\Auth$ (and $\PP$) will act as a client who has complete access to the database. To facilitate this, the data owner first engages in a remote attestation protocol with $\Auth$ along with a secure channel establishment and if it succeeds, she sends the \textit{master} credentials along with the database of client credentials and access policies to $\Auth$ through the established channel. On obtaining these, $\Auth$ uses the SGX seal operation to encrypt and store them.

\subsection{Client authentication}

One of the challenges we need to address is to make sure that only the authorized users 
can query the encrypted database system. For this, we design an authentication method built on top
of an existing DBMS. 

After the database server is started, it is now ready to accept connections from the clients. Here, \name{} adds an authentication mechanism for the clients to authenticate to the $\Auth$ enclave. 
This works as follows.

First, the client proxy verifies that the DBMS has loaded the \textit{correct} code into $\Auth$, by performing the remote attestation (plus secure channel establishment) protocol with $\Auth$ as described in Section \ref{sec:sgx_background}. Let $\sessk$ be the shared secret key obtained after its successful completion.
The client will then authenticate to the $\Auth$ enclave using its credentials, say its password or its SSH key, through the established secure channel. 
On the server side, the I/O layer directs the client authentication requests to the $\CompleteClientAuth()$ function in $\Auth$. $\CompleteClientAuth()$ unseals the  client credentials database and uses it to verify the client credentials. If the client authentication completes successfully, the shared secret key $\sessk$ will be used as the \textit{session key} for the client.

Once the client authentication is completed, the interaction with the client for query processing will be performed by the $\PP$ enclave.
To facilitate this, the I/O layer will now invoke the $\TokenTransfer(\ID, \sessk)$ function in $\Auth$ to transfer the client ``ID'' and $\sessk$ to $\PP$. This transfer will use the secure channel established between these enclaves during the master key transfer. %The $\TokenReceive$ function of $\PP$ will seal and store $\sessk$ with $\ID$ as the additional authentication data during the seal operation.

\subsection{Query execution}
Now we will explain the working of query processing and execution in \name{} for a client which has completed its authentication successfully. 
The design of \name{} permits the use of an unmodified query driver (e.g. JDBC, ODBC, etc.). 

When a client issues a query, the client proxy encrypts the entire query string using the session key $\sessk$ with its $\ID$ included in the additional authenticated data. 
On the server side, the I/O layer directs the client queries to $\PP$. The $\QueryPreProcessing$ function first decrypts the query ciphertext using the session key $\sessk$ for $\ID$. 
Then, it checks whether this client is permitted to run this query. 
%\snote{need to talk more about the policies allowed/implemented/theory 
%of query permissions on dbms. maybe not here but somewhere else}
%\\\snote{what is policy? this paragraph seems to have come up out of nowhere }
Typically, a DBMS allows the DB owners to specify access control policies for the clients.
In \textsf{StealthDB}, we rewrite the access control monitor inside $\PP$. %and the checks can be invoked with the $\QueryControl$ function in $\PP$. 
If the checks are passed, $\QueryPreProcessing$ %runs our version of $\QueryParser$ with the client query $\query$ as input. $\QueryParser$ 
identifies the data values in the query which correspond to the columns in the database using encrypted datatypes using our query parser, and AES-encrypts these data values using the master secret key $\key$. 
%\\\snote{query parse knows the scheme.}
%\\\dnote{I don't understand the comment.}
%\\\snote{when a query arrives, does the query parser need to know the scheme to prepare the new query? 
%i.e. perhaps different value types are encrypted differently (encint vs encstr)?}
The output of this step, $\encquery$, is given to the DBMS for execution.
%\\\snote{add an example here. plaintext query given, in enclave, processed and outputted enc query.}
%\\\dnote{self: ask Alex}

Note that the DBMS is oblivious to the changes made to the query. The \textit{structure} of $\encquery$ is same as that of the query issued by the client. This lets the DBMS use an unmodified query parser to parse this query. But after the query is parsed and a query plan is obtained, we need to augment the DBMS with functions to operate on the encrypted datatypes. We do this as follows.

We first identify the set of \textit{primitive} operators used by the underlying DBMS. Primitive operators are those further-indivisible operators used in query plans:
\begin{itemize}
\item \emph{Arithmetic operators} such as $+, -, \%, *$, etc.
\item \emph{Relational operators} such as $<, >, <=, >=, <>$, etc. 
\item \emph{Logical operators} such as AND, OR, NOT, etc. 
\item \emph{Hash functions} that are used to build some indexes. 
\item \emph{Advanced math functions} such as \emph{sin, cos, tan}, etc. 
\end{itemize} 

\begin{figure}[t]
\centering
\includegraphics[scale=0.4]{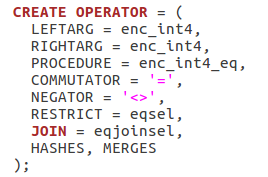}
\caption{Operator $=$ for \textit{enc\_int4}. Here, \textit{enc\_int4\_eq} will call the $\OP$ enclave to decrypt the input, check their equality and output the result.}
\label{fig:encop}
\end{figure}

Traditionally, DBMSs define a functionality for each input datatype tuple supported by a primitive operator.
%\dnote{For example, -- provide Postgres function for + with int and string ---}
\name{} augments these with their functionalities when used with the corresponding encrypted datatypes as in Figure \ref{fig:encop}. 
%Most(?) of the relational DBMSs permit augmenting new datatypes and the operator functionalities for those datatypes. 
Our implementation on Postgres implements primitive operator functionalities over the encrypted datatypes and include them as \textit{extensions}.

For every possible input datatype tuple, we define a function inside the $\OP$ enclave. Suppose that we are given two encrypted data values ($e_1, e_2$) and an operator $\oplus$, the corresponding function inside $\OP$ will perform:  
\begin{enumerate}
\item \textit{decryption} of the inputs $e_1,e_2$ using the master key to get plaintext values $p_1, p_2$,
\item \textit{perform the operator function} to get $p_\out = p_1 \oplus p_2$, 
\item \textit{encrypt} the result $p_\out$ to get a ciphertext $e_\out$ using the master key (if specified by the design).
\end{enumerate} 
The number of inputs and outputs may of course vary depending on operator. 
Moreover, datatype conversions are also allowed in our model. For example, an encrypted integer may be converted to an encrypted string, and so on. 
Overall, we only perform a few basic operations (decrypt, primitive operator, encrypt) during the query execution inside the enclave. 

Finally, once the final result of the query is obtained, $\PP$ re-encrypts the results using the session key $\sessk$ and send them back to the client proxy.

Standard SGX \emph{ocall/ecall} communication mechanism with enclaves is too slow when
many calls are needed. To solve this, we implement an \textit{exit-less} mechanism \cite{eleos} for communicating with $\OP$. In \cite{eleos}, there is always one thread running inside an enclave listening for operator jobs. The DBMS uses our I/O layer to send jobs and receive replies via a communication queue. 
This method greatly improves performance by avoiding context switch for each call to the operator between trusted and untrusted zones, as we discussed earlier in Section \ref{sec:designs}.

There are also other inherent advantages with our design.
\begin{itemize}
\item When a client issues a query only involving unencrypted datatypes, the query processing and execution proceeds in the \textit{native} way and hence with no overheads.
\item A very interesting property is that our design also allows for computations between encrypted datatypes and unencrypted datatypes. The database owner here can also specify that the output of such computations should be encrypted to avoid leaking information about the encrypted inputs.
\item  Since our design implements only the primitive operators, it is easy for us to implement them inside $\OP$ using data-oblivious methods \cite{OSF+16} with a small performance overhead to counter the side-channel attacks of SGX.
\end{itemize}

\subsection{Encrypting indexes}
The indexed columns, unlike the other columns in the table, need extra layers of protection.
When the column is indexed into a B-tree, for example, the structure of the tree reveals the inequalities with respect to the values in the column even though the individual values in the tree are encrypted. 
The inequalities are available even to a snapshot adversary after index creation before any query is made to the database. 
We provide two modifications to reduce this leakage.
First, we re-encrypt the individual values in the column when placing these encrypted values in an index structure. This unlinks the connection between the values in the table and the index. This unlinking is maintained for an adversary obtaining only a snapshot of the table and the indexes. Even for a slightly weaker persistent adversary which does not observe the system during the index creation, the inequalities observed from the index structure can be connected to the table values only when a query accesses the corresponding table row as part of its result. For an adversary persistent throughout the index creation and usage, the security reduces to that provided by order-revealing encryption (ORE) \cite{LW16} on the indexed columns. This change does not incur a performance overhead during the query execution in \name{}.

The second change deals with this leakage on disk. \name{} encrypts every page that is written to the files on disk corresponding to the indexes. 
%\dnote{iterate to precisely specify the adversary type that we are protecting against here}
%We could not do this efficiently by just augmenting the unmodified DBMS with additional components as we have been doing till now.
We do this by encrypting the data right before it is written to the index files on disk, and decrypting the data read from the index files right after it is read from disk.
In our implementation for Postgres, our changes to the codebase involve adding three lines of code to do this task. We create and run a fourth enclave $\IndexOP$ during the DBMS initialization which performs the encryption and decryption of the index data pages. And the three new lines are for retrieving the enclave ID, calling the encryption function inside $\IndexOP$ right before a $\FileWrite()$ of Postgres and for calling the decryption function inside $\IndexOP$ right after a $\FileRead()$.
The key used for these routines is generated and stored by $\IndexOP$, and $\Auth$ attests the correct loading of $\IndexOP$ during the DBMS initialization. 

\subsection{Extensions}\label{sec:extensions}
\paragraph*{Encrypting logs}
\dnote{ does not provide any rationale for not including them in the current implementation}
Some of the log files reveal sensitive information about the queries even for a snapshot adversary on disk \cite{GRS17}.
We can protect against an adversary accessing disk by encrypting the log files on disk in a way similar to our encryption of index files on disk.
Perhaps, one could ask why we do not encrypt every page written to disk, not just indexes and logs. But the individual data items in the tables are already encrypted and we get no concrete security improvements by encrypting the individual disk pages containing those data items.

\paragraph*{Key management}
In the current implementation, we use a single master key $\key$ to encrypt all the data values. $\key$ is sealed and stored on the disk by $\PP$ or $\OP$ enclave when obtained from $\Auth$. If and when the system is restarted, the enclaves are created again and a valid $\PP$ or $\OP$ enclave can unseal the corresponding sealed components to obtain $\key$. During this process, the AES-GCM encryption used in the SGX sealing provides confidentiality and integrity for the sealed component of $\key$ against any adversary. Also, when replicating the database across multiple machines, we can let the $\Auth$ in one of the machines to generate $\key$ and do a remote attestation to transfer it to the $\Auth$ enclaves in the other machines.

\section{Security evaluation}\label{sec:security}
The tradeoff between security, functionality, performance and the intrusiveness level decided by our design results in the leakage profile that we explain in this section. %For comparison, our leakage profile matches the leakage profile of the strongest version of \cite{cipherbase}, which is the state-of-art for the encrypted database systems which offer reasonable performance and high scalability for transactional workloads.

First, we will discuss the effect of the SGX side-channel attacks on \name{}. SGX is subject to various side-channel attacks as described in Section \ref{sec:sgx_background}. The side-channel due to the  application's page-level access pattern is a significant one and it is up to the application developer to design data-independent memory accesses for the application data to be secure.
Our design addresses this side-channel by performing only primitive operations inside an enclave (Sections \ref{sec:designs} and \ref{sec:architecture}) and by using oblivious operators \cite{OSF+16} for these primitive operations. 
We obviate the other software side-channels (except the cache-based ones) by simplifying the code inside the enclaves; running the primitive operations obliviously prevents these side-channels. The cache-based side channels \cite{foreshadow,cachequote} though, are inherent to the x86 architecture and requires patching from Intel. (Also, these are instances of active attacks, which in general StealthDB does not protect against).

Now, let us discuss the leakage profile of \name{}. As mentioned in our threat model in Section~\ref{sec:threatmodel}, \name{} protects against semi-honest or passive adversaries. It does not provide integrity guarantees to the clients on the correctness of the query results. Neither does it provide confidentiality guarantees against an actively malicious adversary with side-information on the plaintext values encrypted in $\DB$. %This would be an interesting follow-up to \name{}. 
We will first detail the leakage profile of \name{} for different variants of semi-honest adversaries and through a series of security claims we will argue that \name{} does not leak any more information than what is part of the leakage profile. Our evaluation is with respect to the architecture we propose, and hence independent on the specific underlying DBMS engine.

\subsection{Leakage profile} \label{sec:leakprof}
\name{} encrypts the individual data items, rather than an entire column or table at once, and hence this mandates a thorough leakage profiling. We classify the admissible adversaries as in \cite{sokencdb} and quantify leakage profiles during the high level operations, $\Init$ and $\Query$, of a DBMS for those adversaries. $\Init$ involves loading the database in the untrusted server to be ready for querying, and $\Query$ involves the client querying the database to get the required results. Note that a $\query$ in \name{} can involve any operator supported by the underlying DBMS (for eg., relational, arithmetic and logical operators for a transactional DBMS).
%While doing so, we compare how \name{}'s leakage profile fares against other encrypted DBMSs offering complete functionality.

%the database schema is revealed to the adversary. In general, \name{} offers a black-box access for primitive computations over encrypted data items to the DBMS.
We analyze the security of \name{} against passive or semi-honest adversaries. We further classify the adversaries into snapshot and persistent adversaries. A snapshot adversary gets a snapshot to the memory of the system whereas a persistent adversary observes the memory of the system throughout its execution. We motivate these adversarial types in Section \ref{sec:threatmodel}. A formal security definition is provided in Appendix \ref{sec:encdbdef}.

% OPE-based constructions do not provide security against snapshot attackers too \cite{??}. We either match or improve the leakage profile from prior designs evaluated for complete DBMS \cite{cipherbase,cryptdb}. We let an adversary know details of DBMS execution and hence will know how to parse
%
Let $\DB$ denote the database that we try to securely operate. $\DB$ includes all the data structures used by a database (for eg., tables, indexes, views, foreign tables) along with their contents. We will now define the leakage entities to understand the security of \name{}. 
To understand the security of our system, we study the leakage profile during different phases of database execution: during the steady state and query execution.
%Our security argument for \name{} will follow along the lines of a simulation paradigm. In such a security proof, one first defines the set of admissible adversaries. Then a polynomial-time simulator $\Sim$ is constructed which, with only the leakage entities as input, produces a view (``ideal view'') that is indistinguishable for any admissible adversary from the view (``real view'') obtained through a real execution of the system. This will prove that the upper bound on the information leaked by the system to the adversary are the inputs used by the simulator to produce the ideal view. 
The leakage entities of interest to \name{} are as follows:
\begin{itemize}
\item Let $\cS_t$ indicate the \textit{shape} of the database at time\footnote{``Time'' $t$ refers to the epoch at which the data-structure is observed or collected from the system} $t \geq 0$ which includes 
\begin{itemize}
\item the database schema, 
\item the shape of the tables and (database) views i.e., the number of rows and columns in the tables and views,
\item the shape of the indexes (for eg. the shape of a B-tree index reveals the number of keys in each internal node of the tree).
\end{itemize}
More importantly, $\cS_t$ does not include the contents of any of the data structures in the database. This entity varies with time depending on the queries run on $\DB$.
\item Let $\cQ$ denote the leakage associated with a query execution. In \name{}, $\cQ$ is upper bounded by the union of the plaintext outputs of the $\OP$ enclave invocations.
\item Let $\cM_t$ denote the leakage associated with the logs and the miscellaneous data structures maintained by a DBMS at time $t$ to aid in its operations (including various profiling activities and recovery from unexpected failures).
\end{itemize}
In \name{}, the entities $\cQ$ and $\cM$ are dependent on the underlying DBMS that \name{} builds on. 
%Some of the data structures (queries and logs) reveal more information than desired for security. 
In Section \ref{sec:leakage}, we discuss the information that can be inferred from $\cS$, $\cQ$ and $\cM$ for some real-world data structures and queries. 

Note that $\cS$, $\cQ$ and $\cM$ are leakages with respect to $\DB$. We now define the leakage entity $\cq$ with respect to a $\query$. In \name{}, before the query is executed (after output by $\PP$), the query structure is revealed but not the constants in the query which are encrypted with the semantically secure encryption. With $\cQ$ being the leakage during the execution of this query, the total leakage of a client query to the server is upper-bounded by the union of $\cq$ and $\cQ$ for this query.
\begin{itemize}
\item Let $\cq$ indicate the leakage about the $\query$ before the DBMS begins processing it.
\end{itemize}
Typically, $\cq$ will be a subset of the $\DB$-based leakages. In a real-world DBMS, $\cq$ might just be a subset of $\{\cM_t\}$ since the details about input queries are usually logged and checkpointed.

We will now argue the leakage profile of \name{} during different phases of its execution. All the following claims rely on the fact that no information (other than its length) about the key $\key$ used to encrypt the data is revealed to an adversary (Claim \ref{cl:key}). We would rely on the following security properties:
\begin{enumerate}
\item Remote and local attestation provided by SGX are secure according to Section~\ref{sec:sgx_background}.
\item The confidentiality of the intermediate values of the computation and the integrity of the computation from SGX.
\item The confidentiality and integrity provided by the secure channels established. %-- use public-key encryption like ElGamal and AES-GCM.
\item The confidentiality and integrity of the SGX sealing procedure.
\end{enumerate}

\paragraph*{$\Init$ phase}
\name{} only leaks the initial shape $\cS_0$ during the $\Init$ phase. This is better than the OPE or ORE based designs \cite{cryptdb,seabed} which leak the `<' relation between all the values in the OPE/ORE encrypted columns.
\begin{claim}
After the completion of $\Init$ and before any call to $\Query$ is made, \name{} leaks at most $\cS_0$.
\end{claim}
The high-level idea of the correctness of this claim is as follows.
$\Sim$ obtains $\cS_0$ from the leakage oracle $\cL$ and outputs encryption of zeros according the shape $\cS$ as $\EDB$. An adversary $\Adv_2$ that distinguishes the simulated $\EDB$ from a real $\EDB$ will break the semantic security of the encryption scheme.
%\paragraph*{Snapshot adversary}
%An adversary which obtains a snapshot access to the memory learns the index structure (for e.g., the structure of the B-tree for a B-tree indexed column), for the part of the index that resides in the memory during the adversarial access. Though the individual data items remain encrypted, the index structure might reveal some non-trivial information about the data items.
%%to enable an \textit{unmodified} DBMS to process queries \textit{efficiently} using the indexes built. 
%But, since we encrypt the index pages before they are written disk, we leak no information about the entire dataset to an adversary which obtains a snapshot access of the disk.
\paragraph*{Query phase}
We will first argue the leakage of \name{} for adversaries which obtain snapshot access to the memory. A snapshot adversary in \name{} learns at most the shape $\cS$ and the leakage $\cM$ due to the miscellaneous information maintained at the time of the snapshot. $\cM$ is further upper-bounded by the union of $\cQ$ from the queries executed recently. More formally, we have the following claim. The correctness arguments for the claims in this section are in the Appendix.
%We will explain what these mean for a real-world DBMS in Section \ref{sec:leakage}.
\begin{claim}\label{cl:snapshot}
Consider a polynomial-time snapshot adversary on \name{} obtaining the snapshot at time $t$. Let $t'\leq t$ be the latest time epoch before $t$ for which the logs and miscellaneous data structures remain in memory and not written to disk. The adversary learns at most $\cS_{t'}$ of the $\DB$ being operated and $\cQ$ of the queries executed between $t'$ and $t$. If the log items are encrypted in memory and assuming that the size of logs do not reveal sensitive information, the adversary learns at most $\cS_t$.
\end{claim}

We will now argue the leakage for a persistent adversary. A persistent adversary in \name{} learns the plaintext outputs of the $\OP$ enclave invocations throughout its observance.
%, and the parts of $\DB$ being accessed during the query executions. ---IMPORTANT--- 
%For example, when a comparison is made between two encrypted data values of an indexed column, the resulting branch of the B-tree being accessed reveals the result of the comparison. 
More formally, we have the following claim.
\begin{claim}\label{cl:persistent}
A polynomial-time semi-honest adversary that has persistent access to the memory during the \name{} execution on a $\DB$ learns at most the shape $\{\cS_t\}_{t \geq 0}$ of $\DB$ and the query-execution associated leakage $\cQ$ for all the queries executed, where $\cQ$ is the union of the plaintext outputs of $\OP$ invocations during the execution of the query.
\end{claim}
Note that this claim implies that the miscellaneous data structures $\cM$ maintained or the parts of $\DB$ accessed during query execution do not leak more information than $\{\cS_t\}$ and $\{\cQ\}$ to a persistent adversary.

\section{Implementation and Performance}\label{sec:implementation}

\subsection{Implementation details}

\begin{figure}[h]
 \centering
  \includegraphics[scale=0.4]{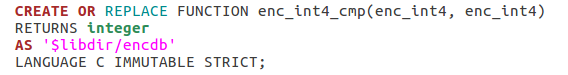}
 \caption{Example of a new function definition in stealthdb.sql}
 \label{fig:fndefeg}
 \end{figure}

We implement StealthDB in C and C++ on top of Postgres 9.6 as an extension that loads new SQL functions, encrypted data types and operators and index support methods for the encrypted datatypes. 
\ignore{
\begin{figure}[h!]
 \centering
 \hspace*{-0.5cm}\includegraphics[scale=0.6]{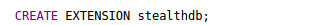}
 \end{figure}
}
The command \textit{CREATE EXTENSION stealthdb}  loads the files \textit{stealthdb.so} (the main library),  \textit{enclave\_stealthdb.so} (part of the code which is executed in enclaves),  \textit{stealthdb.control} (the version control file), \textit{stealthdb.sql} (definitions of new defined functions) into the system. 
For instance, the function \textit{enc\_int4\_cmp} in Figure \ref{fig:fndefeg} compares two \textit{enc\_int4} values and returns \{-1, 0, 1\}.  
\begin{figure}[h]
 \centering
   \hspace*{0.1cm}\includegraphics[scale=0.4]{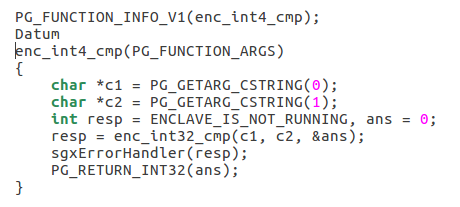}
 \caption{Example of new defined function implementation in stealthdb.c}
 \label{fig:fnimpleg}
 \end{figure}
The function \textit{enc\_int4\_cmp} in Figure \ref{fig:fnimpleg} is executed in an enclave.
We implement our query \textit{pre-parser} in the $\PP$ enclave on the server side to encrypt the data values in queries and this design helps in avoiding changes to the client JDBC or ODBC drivers of the system. 
Our approach can be extended to other SQL-like database using user-defined functions. 
Though database systems like MySQL do not allow creating independent extensions like Postgres to include our changes, these changes are not intrusive and completely independent of the improvements to the core database operations. 
To protect against the side-channel attacks on SGX, we make every operation inside an enclave oblivious by leveraging AES-NI and CMOV instructions. 
The source code of Postgres 9.6 has about 700k lines of code while \name{} has about 5k lines of code with 1.5k lines run in enclaves.  

\subsection{Performance evaluation}
To measure StealthDB's performance, we use an Intel Xeon E3 3.60 GHz server
with 8 cores and 16 GB of RAM. In our experiments, we measure the throughput and
latency of \name{} using the TPC-C trace and compare the results with an unmodified Postgres 9.6 which works with unencrypted data. 
%We choose to evaluate an option where IDs are unencrypted because these are auto-generated at the tables and do not directly
%contain any sensitive data. Their relationship structure may leak some information, but the performance improvements 
%are significant to consider. 
%\anote{we should tell here why we choose this option?}
The results were obtained by averaging multiple 1000 second runs with check-pointing turned off. We ran our experiments with the number of clients varying from 1 to 10 and with a single-threaded enclave used by all the client connections. The number of clients can be further increased if a multi-threaded enclave is used. Our first set of experiments leave the IDs in the TPC-C tables (e.g. w\_id, o\_w\_id, etc.) unencrypted. The tested database includes nine tables with about 10 million rows in total. This is about 2GB of unecrypted data and when encrypted for \name{} gives an encrypted database of size 7GB.

\begin{figure}[t]
 \centering
  \hspace*{-0.25cm}\includegraphics[width=0.5\textwidth]{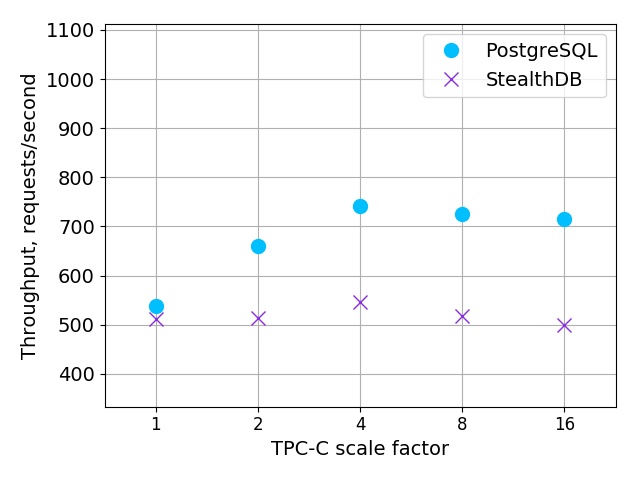}
 \caption{TPC-C benchmarking throughput for running under Postgres and StealthDB with different scale factors}
\label{throughput}
 \end{figure}

\paragraph*{Throughput} Figure \ref{throughput} shows the throughput for the TPC-C benchmarking for different scale factors. \name{} incurs an 4.7\% overhead over the unmodified Postgres for the scale factor $W=1$ and around 30\% overhead for $W=16$. 
%This result is competitive with the best existing encrypted databases and sufficient for many real-world transactional systems. 
This is sufficient for many real-world transactional systems for the security advantages.

\paragraph*{Latency} We measure the end-to-end TPC-C transaction latency for \name{} with the scale factor $W=16$. This includes the time for our query pre-parser. 

%Figure \ref{latency} shows CDF graphs for each type of TPC-C request.
%\begin{figure}[h]
%\centering
%\begin{subfigure}{.5\textwidth}
%%  \centering
% \hspace*{-0.45cm}\includegraphics[width=.55\textwidth]{table7.jpeg}
%\end{subfigure}%
%\begin{subfigure}{.5\textwidth}
%%  \centering
%  \hspace*{-4.35cm}\includegraphics[width=.55\textwidth]{table8.jpeg}
%\end{subfigure}
%\begin{subfigure}{.5\textwidth}
%%  \centering
%   \hspace*{-0.55cm}\includegraphics[width=.55\textwidth]{table9.jpeg}
%\end{subfigure}%
%\begin{subfigure}{.5\textwidth}
%%  \centering
%  \hspace*{-4.5cm}\includegraphics[width=.55\textwidth]{table10.jpeg}
%  \label{fig:sub2}
%\end{subfigure}
%\label{fig:test}
%\begin{subfigure}{.5\textwidth}
%%  \centering
%     \hspace*{1.5cm}\includegraphics[width=.55\textwidth]{table11.jpeg}
%\end{subfigure}
%\caption{Cumulative distribution functions for~TPC-C requests under Postgres and StealthDB. For all graphs the left line (solid) represents Postgres, the second line - StealthDB with unencrypted IDs, the right - StealthDB with encrypted IDs.}
%\label{latency}
%\end{figure}

\begin{figure}[h]
 \centering
     \hspace*{-0.3cm}\includegraphics[width=0.5\textwidth]{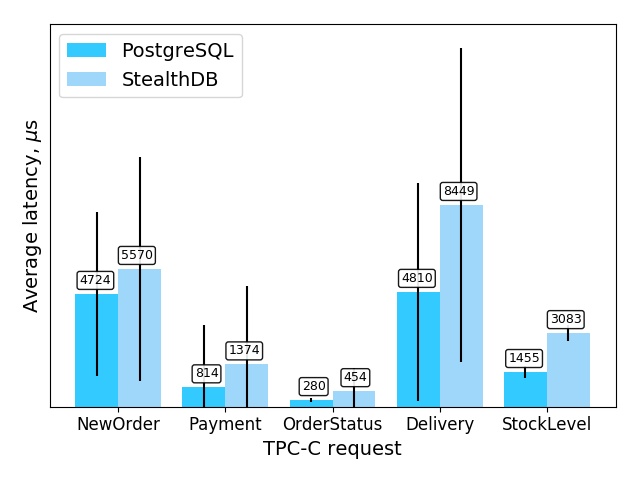}
 \caption{Average latency and standard deviation for TPC-C requests under Postgres and StealthDB.}
\label{avg_lat}
 \end{figure}

Table~\ref{median_lat} and Figure~\ref{avg_lat} compare the median and average latency for StealthDB with the unmodified Postgres. The 90th percentile of the latency of StealthDB system is 7.2 milliseconds which results in a 22\% overhead over the unmodified version.

\begin{table}[h]
\begin{center}
\small
\begin{tabular}[ht]{ | m{2.2cm} | m{1.0cm}| m{1.4cm} |} 
\hline
& \tt Median & \tt 90th \tt  percentile  \\ 
\hline
\tt PostgreSQL & \tt 1.6 & \tt 5.9  \\ 
\hline
\tt StealthDB & \tt 2.8 & \tt 7.2 \\ 
%\hline
%StealthDB, full encryption & 15.2 & 26.1 \\ 
\hline
\end{tabular}
\end{center}
 \caption{Latency statistics of TPC-C requests, ms}
\label{median_lat}
\end{table}

We also test the performance of \name{} when the IDs are encrypted with AES-CTR. That results in about %40\% storage overhead, 
3x throughput decrease over \name{} with unencrypted IDs. And the latency is 3.6 times of that of the version with unencrypted IDs.
The IDs in the TPC-C tables are just counters, hence encrypting them do not offer any concrete security advantages.

%\subsection{Performance comparisons}
%The only other TPC-C benchmark results on encrypted data that we are aware of were provided for the CryptDB [???] and Cipherbase [???] systems. %Moreover, these systems used modified and incomplete TPC-C requests. 

%\anote{need some conclusion?}

\section{Related Work}\label{sec:relatedworks}
%\dnote{The related work section can be substantially improved to explain the novelty of StealthDB. The current statements appear vague. E.g., "We explore this design in Section 4.2 and conclude that it’s not ideal when working with SGX."}
This section builds on the comparisons from the introduction.
The work most similar to ours is Cipherbase~\cite{cipherbase}. 
But the trusted on-premise key loading phase for every FPGA device, and cloud operator controlled ``shell'' monitor \cite{awsf1shell} inside an FPGA make FPGAs unsuitable for being used as a \textit{trusted hardware} in the cloud. 
In terms of performance, \cite{cipherbase} achieves about 10\% better throughput than ours, but they skip two TPC-C transactions in their evaluation. Our evaluation with the complete TPC-C benchmark finds that these two transactions have the highest latency overheads. Similar bottlenecks are expected for Cipherbase with FPGAs. And, as expected, we achieve much lower latency ($4\times$) over the FPGA implementation.
%The end-to-end security of these have not been studied yet.
%On the other hand, though SGX comes with its own bag of limitations, we are able to design \name{} to get around these limitations. 
%We also 
%build an overall design for an encrypted data, discussing authentication methods, 
%and extensively evaluate our system. 
TrustedDB~\cite{trusteddb} uses the IBM secure co-processor to perform operations, but with large portions of the DBMS engine executed 
inside the trusted zone. The IBM co-processor incurs high overheads for transactional workloads and also, this design is not suitable for SGX for both security and performance reasons as we discussed in Section~\ref{sec:designs}. 
%We explore this design in Section~\ref{sec:designs} and conclude that it's not ideal when working with SGX. 

CryptDB~\cite{cryptdb} uses a hybrid of encryption schemes 
to support subset of SQL functionality. Their underlying large leakage profiles
often result in data compromise~\cite{inference,breakingmylar}. Performance-wise, \cite{cryptdb} achieves a similar throughput decrease as ours, but only when evaluated with the individual queries from the TPC-C transactions over a $20\times$ smaller dataset.
Arx replaces OPE scheme with a special garbled-circuit based searching method~\cite{arx}. 
Garbled circuits however introduce large computational and storage overheads.

%\dnote{more concrete comparisons for the following paragraph..}
A few works studied how to build versions of encrypted databases 
with SGX. 
VC3 system proposes an architecture for analytical MapReduce jobs in cloud settings~\cite{VC3}. 
Opaque studies how to leverage SGX to secure distributed analytical workloads in Spark systems~\cite{opaque}. 
A concurrent work of ours, ObliDB \cite{oblidb}, obtains an oblivious database supporting both transactional and analytical workloads. But, their solution involves extensive changes to the underlying DBMS engine, and does not scale well for transactional workloads.
Another concurrent work, EnclaveDB \cite{enclavedb}, provides strong security guarantees against persistent and active adversaries. However, this is achieved by placing larger components of DBMS inside enclaves assuming the existence of large enclaves, in the order of gigabytes, which is much greater than the 128 MB available today.\footnote{It is an open question to achieve larger enclaves efficiently while providing security against physical attacks. SGX enclaves use Merkle-trees for integrity which adds logarithmic overhead to every access.} They also ignore the access pattern and other side-channel attacks. In summary, \cite{enclavedb} focuses on a different design space assuming how future trusted hardware designs may look, while our work focuses on building encrypted database from standard trusted hardware available today.
%1) EnclaveDB requires more intrusive DB changes. 

HardIDX~\cite{hardidx} investigates how to perform range queries obliviously over B+ tree indexes inside an enclave, leaking only the parts of the database accessed per query. But, they only consider a static database, and the client should generate the full B+ tree index locally and store it in the server only for the querying. We can incorporate their ideas in \name{} if we were to only support static databases and powerful clients. Also, \cite{hardidx} just prototypes index searches, whereas we architecture and build a complete encrypted database system. 

A number of works study how to load unmodified applications into enclaves~\cite{haven,scone,graphene,ryoan}. These approaches work well for applications that process small data sizes, but do not scale well to larger workloads due to 
SGX limitations. Also, increasing the complexity of the codebase inside the enclaves aggravates the security risks associated with SGX \cite{LJJ+17}.

OSPIR-OXT \cite{ospir1,ospir2,ospir3}, SisoSPIR \cite{sisospir} and BLIND SEER \cite{blindseer} build encrypted database systems from scratch with provable security guarantees for a subset of functionality based on different cryptography tools. There are also multitude of other works which provide improvements over security or specific functionalities of a database, but they are not implemented or integrable with a mature DBMS. A recent systematization work by Fuller et al.~\cite{sokencdb} provides are great summary of the state-of-art research in encrypted database systems. 
Fully homomorphic encryption \cite{Gen09} is another powerful cryptographic primitive which enables an untrusted user to perform arbitrary computations on encrypted data without learning any information about the underlying data. But the current constructs for doing this are very far from being practical \cite{helib}.
In general, while theoretical security of systems built based on cryptographic methods can be high, the \textit{real-world} security of the system relies on the multitude of factors: correct implementations of non-trivial crypto algorithms, meta-data contents, information in log files, %database structure and stored relationships in data-structures, 
etc. Hence, it is not possible to argue their security just from the security of the crypto protocols used.

\section{Conclusion}\label{sec:conclusion}
%Encrypted databases are important when one tries to protect sensitive data from unauthorized users, attackers and administrators in cloud settings. 
\name{} offers a scalable encrypted cloud database system with full SQL query support with a modest 30\% throughput decrease and $\approx 1$ ms latency increase while providing strong end-to-end security guarantees. 
\name{} can be implemented in any newer generation Intel CPUs. Supporting analytical workloads, reducing the leakage profile and protecting against active adversaries (i.e., providing integrity to the system) while maintaining our design principles are interesting open questions in this space.
The source code of our implementation is also open-sourced. %available at \url{https://github.com/cryptograph/stealthdb}. 

\bibliography{main}
\bibliographystyle{abbrv}
%\newpage
\appendix
\subsubsection*{Acknowledgements}
S.G. was supported by grants from NSERC and University of Waterloo.
The authors would like to thank the reviewers and the shepherd for their great comments and suggestions on improving the quality of the paper.
D.V. would also like to thank Hemant Saxena for discussions on the internals of Postgres..
\section{Security addendum}
\subsection{Formal definition of security}\label{sec:encdbdef}
Figure \ref{def:encdb} provides the formal simulation security definition for an encrypted database system using trusted hardware definition. This definition is inspired by \cite{iron} who define simulation security for functional encryption using trusted hardware $\HW$.
\begin{figure*}[t]
%\begin{table*}[t]\label{def:encdb}
\begin{center}
%\centering
\begin{tabular}{ll}
${\Real_{\EncDB}}(1^\secp):$ \hspace{20mm} & ${\Ideal_{\EncDB}}(1^\secp):$ \\[5pt]
$(K, \EDB) \gets \Init(1^\secp, \DB)$ & $\EDB \gets \Sim^{\cL}(1^\secp)$\\[5pt]
$\encres \gets \Query^{\HW(\cdot)}(\EDB, \encquery)$ &	$\encres \gets \Query^{\Sim^{\cL(\cdot)}(\cdot)}(\EDB, \encquery)$\\[5pt]
%$\ct \gets \feenc(\mpk, \msg)$ & $\ct \gets \cS^{U_\msg(\cdot)}(1^\secp, 1^{|\msg|})$\\[5pt]
%$\alpha \gets \cA^{\fekeygen(\msk, \cdot), \HW, \ATH(\cdot)} (\mpk, \ct)$ & $\alpha \gets \cA^{\cS^{U_\msg(\cdot)}(\cdot)} (\mpk, \ct)$\\[5pt]
%Output $(\msg, \alpha)$ &	Output $(\msg, \alpha)$\\[5pt]
\end{tabular}
\end{center}
\caption{Security definition for an encrypted database system using trusted hardware.}
%\begin{flushleft}
%\end{flushleft}
\label{def:encdb}
%\end{table*}
\end{figure*}
An $\EncDB$ construction is secure if, for all admissible adversaries, there exists an efficient $\Sim$ such that:
\[
\left| \Pr[\Adv(\Real_{\EncDB}) = 1] - \Pr[\Adv(\Ideal_{\EncDB}) = 1] \right| < \negl(\secp)
\] 
where $\Adv = (\Adv_1,\Adv_2)$. $\Adv_1$ runs the $\Real$ or the $\Ideal$ experiment, whereas $\Adv_2$ obtains information about the experiment from $\Adv_1$ depending on the adversarial type being studied and produces the output 0 or 1. A snapshot $\Adv_2$ obtains a snapshot of the system, when desired, from $\Adv_1$, whereas a persistent $\Adv_2$ completely observes the $\EncDB$ system while $\Adv_1$ is running the experiment. $\Adv_1$ is tasked with just running the $\EncDB$ system; a semi-honest $\Adv_1$ will run as per the specifications, and an actively malicious $\Adv_1$ will run the system as desired to maximize the information obtained by $\Adv_2$.
The access to $\HW$ is treated as an oracle as in \cite{iron} and $\Sim$ simulates the oracle in the $\Ideal$ experiment. The $\HW$ oracle provides interfaces to the enclaves used (in \name{}, they are $\Auth(), \PP(\encquery)$ and $\OP(\{\inp\}, \op)$). When $\Query$ is invoked on a $\query$, $\Sim$ will obtain the leakage $\cQ$ corresponding to a query from $\cL$.
 
\subsection{Security of $\key$ during \name{} execution}
%The primitives which we serve as our security building blocks are Intel SGX and AES. 
\textit{Outline.} We will argue here that no information about the master key $\key$ is revealed; also that only the \textit{permitted} clients can make the DBMS execute queries. This will be a precursor to the leakage profile analysis in Section \ref{sec:leakprof} %then use these to argue the semantic security of the data values during the initialization of the encrypted database before the queries are executed; finally, use all these to argue the leakage during the query execution phase.

\begin{claim}\label{cl:key}
The confidentiality and integrity of the master key $\key$ is ensured throughout the \name{} execution.
\end{claim}
The database owner forms the \textit{root of trust} as in Figure \ref{fig:authprotocol}. The owner is involved in a remote attestation protocol with $\Auth$ to check the correctness of the code and the constants loaded into $\Auth$ against the publicly available \textit{expected measurement} of $\Auth$.
(The constants loaded into $\Auth$ include the expected measurements of $\PP$ and $\OP$).
The master credentials for the database is transferred to a valid $\Auth$. And, the security of SGX remote attestation guarantees the validity of $\Auth$. 
From this point, the trust is transferred to $\Auth$. $\Auth$ generates the master key $\key$.
%We will now argue that no information about $\key$ can be learnt by an adversary (other than its length).

The master $\key$ is then transferred to the other enclaves $\PP$ and $\OP$ by $\Auth$ through the secure channels established on top of local attestation. The security of local attestation ensures that $\Auth$ establishes secure channels with only those $\PP$ and $\OP$ whose measurements match the \textit{expected} hardcoded ones. Hence, $\key$ is transferred only to the correct instances of $\PP$ and $\OP$. 
Here, the confidentiality and integrity provided by the secure channel ensure that no information about $\key$ except its length is leaked to an adversary during the transfers.
%Here, the confidentiality provided by the public key cryptography used in the secure channel establishment (on top of the authenticity from attestation) and the confidentiality and integrity of AES-GCM ensure that no information about $\key$ except its length is leaked to an adversary during the transfers.

Now, there are only two more operations which involve $\key$. First, when $\key$ is used to AES encrypt and decrypt data values, the SGX security guarantees combined with the use of a data-oblivious implementation of the AES-NI instructions ensure that no intermediate values about $\key$ are leaked.
Finally, $\key$ is also sealed and stored on the disk for later retrieval. Here, the SGX sealing process provides confidentiality and integrity to $\key$.

\begin{claim}\label{cl:permitquery}
During the query execution phase, a query which reaches the DBMS for execution satisfies the access control policies for the client requesting the query.
\end{claim}
%We defer the following proofs to our arXiv version due to space restrictions.

The security of remote attestation also ensures that the database owner transfers the client credentials database only to a valid $\Auth$. 
When a client proxy initiates a connection with the DBMS, a valid $\Auth$ establishes a session with the client only if the client has valid credentials.
Next, $\Auth$ transfers the session key $\sessk$ (shared with the client) only to a valid $\PP$. This is ensured by the security of local attestation.
Now, when the client issues a query, the I/O layer relays it to $\PP$ and $\PP$ parses the query and proceeds only if the query satisfies the access policies of this client.
Since there is no other interface for the client to issue a query to the semi-honest DBMS, \name{} ensures that the semi-honest DBMS only executes a query from a valid client satisfying the access policies provided by the database owner.

\subsection{Correctness of Claim \ref{cl:snapshot}}
$\Adv_2$ would query the snapshot of the system at time $t$. $\Sim$ sets up $\EDB$ as encryption of zeros of arbitrary shape $\cS_0$ and answers the $\OP$ queries arbitrarily till $t'$. At time $t'$, $\Sim$ obtains $\cS_{t'}$ from $\cL$ and rewrites $\EDB$ with encryption of zeros according to $\cS_{t'}$. For each query run between $t'$ and $t$, $\Sim$ obtains $\cQ$ from the oracle $\cL$ and answers the $\OP$ queries accordingly. This way, the execution of the $\Real$ and $\Ideal$ experiments and the corresponding shapes of $\EDB$ are consistent at time $t$ assuming a deterministic order of execution for $\EDB$.

We will now argue that the $\Real$ and the $\Ideal$ experiments are indistinguishable. When $\Adv_2$ obtains the snapshot of the system at time $t$, it obtains $\EDB$ along with the logs and miscellaneous data structures maintained at time $t$. Given that the shape of $\EDB$ is consistent between the two experiments at time $t$, semantic security ensures that a real $\EDB$ is indistinguishable from the encryption of zeros. Logs, etc. for queries before time $t'$ are encrypted and written to disk. Hence, they do not reveal any information about the data items in $\DB$. The logs maintained in between $t'$ and $t$ are also consistent between the two experiments and are consistent.

If the logs and the other data structures are encrypted in memory, $\Sim$ can behave arbitrarily till $t$ and just rewrite $\EDB$ according to $\cS_t$ at time $t$. Following the assumption that the size of logs do not reveal sensitive information, the $\Real$ and the $\Ideal$ experiments are indistinguishable to $\Adv_2$.

\subsection{Correctness of Claim \ref{cl:persistent}}
We again give the high-level idea here.
During $\Init$, $\Sim$ obtains the shape $\cS$ from the leakage oracle $\cL$ and encrypts zeros as $\EDB$ according to $\cS$. This $\EDB$ is indistinguishable from a real $\EDB$ by the semantic security of the encryption scheme. Further, during the execution of $\Query$, the values in $\DB$ are only used inside the $\OP$ enclave. With a deterministic execution of $\EDB$, $\Sim$ uses $\cQ$ obtained from $\cL$ to answer the plaintext outputs. For the encrypted outputs, $\Sim$ produces encryption of zeros as $\OP$ output and this is again indistinguishable from the encryption of the real values by the semantic security of the encryption scheme.

\section{Concrete leakage profiles} \label{sec:leakage}
The discussion above provided an upper bound on the leakage in terms of abstract leakage entities. The definition of the shape $\cS$ is concrete from the definition. But, $\cQ$ and $\cM$ depend on the underlying DBMS that \name{} builds on. We will now concretize this for the different operations performed on encrypted data.%$\cM$ is typically  
\begin{itemize}
\item \textit{Arithmetic operations}: Some examples of arithmetic operators include +, -, \%, * and advanced ones like sin, cos, log. For these operators, we provide the same security as a fully-homomorphic encryption (FHE) on the computation performed on individually encrypted data items. As in FHE, \name{} does not reveal any information to a semi-honest adversary about the intermediate values of an arithmetic computation involving encrypted inputs and outputs, other than their length (as multiples of 128 for AES). Consider a simple example query from a TPC-C transaction: \textsf{update table\_warehouse   set w\_ytd = w\_ytd + constant  where w\_id = constant2}. 
%$\cQ$ for this query  
\name{} reveals no information about the values in the column w\_ytd during the execution of this query.
\item \textit{String operations}: String operations like substring and wildcards have no leakage, other than the length of inputs and outputs (up to a multiple of 128), with them being encrypted.
\item \textit{Relational operations}: A real-world DBMS uses indexes to perform the relational operations like comparisons and joins efficiently. The $\cQ$ for a query using an index, say a B-tree, includes the comparison results of the parts of the B-tree explored by the query. As the values in the index are re-encrypted versions of the values in the table, the comparison results are useful only when the corresponding values are accessed in the table. When a row becomes part of query results, an adversary can link it to the corresponding value in the index. From this, it can use the $\OP$ output history to obtain the comparison results between the indexed value in this row with the indexed values from the other accessed rows. %by linking them to the $\OP$ outputs during the index traversal. 
Hence, the information revealed by $\cQ$ in \name{} is the comparison results for indexed values in the rows accessed by the queries. In the worst case, our leakage against persistent adversaries reduces to the guarantees provided by ORE for the parts of the indexes explored by the queries.

There is also a non-trivial information leakage to a persistent adversary that only has access disk, and not memory. The index pages on disk that are modified during checkpointing reveal some inequalities within the data being inserted or modified. In Postgres, for instance, the index file stores data as 8 KB pages. When a new value is inserted into the table, only the pages that need to be changed are marked as dirty in the memory and eventually changed on disk. 

For any other DBMS, the precise information revealed by $\cQ$ varies based on its query execution and log maintenance procedures. 
\end{itemize}
%But we believe that a practical DBMS would attempt to minimize the data being accessed to process a query, and hence any passive adversary is restricted to learning access pattern leakages only for the data \textit{relevant} to the query. Adversary learns no information about the data \textit{irrelevant} to the query, since the constants in the query are encrypted.
%

%\dnote{self: read Opaque to see if we can provide integrity using their techniques}
%\snote{refer to the discussion section for integrity? we will need a paragraph on this at least theoretical follow-ups}

%\snote{there is no need to discuss our shortcomings here. move to discussion/extensions. every leakage should follow by theoretical/possible extensions to protect against it. references.}

%Finally, \name{} also provides semantic security to the whole database during a hypothetical \textit{steady state}. That is, even when the DBMS is up and running, when no queries are being executed at an instant and when the changes made by the previous queries are already written to disk (checkpointed), semantic security is provided to all the data. %\dnote{should we even say this? will a server ever be idle?!}
%%\dnote{we say that we implement log encryption, so we have to atleast mention an estimate for the numbers with it.}

\dnote{can we provide ``simple'' suggestions for the DBMS designers to reduce the leakages, or in general improve security? so that systems like ours can be built on top to provide ``full'' security..\\
1) providing a layer to modify data before it is written to disk?! -- not sure if this is one, but something along these lines, that would have made our design simpler, and achieve stronger security guarantees.}

%\dnote{With every line being picked and highlighted from CryptDB by the papers which attacked it.. makes me fear/responsible for every line that we put in this paper and the tone and the precision of those lines...}
%

\end{document}